%% file: main.tex
\documentclass[longbibliography,onecolumn,10pt,a4paper,sort&compress,aps,pra,showpacs,superscriptaddress]{revtex4-2}

\usepackage{lmodern}
\usepackage{amsmath,amssymb,mathtools}
\usepackage{graphicx}% Include figure files
\usepackage{dcolumn}% Align table columns on decimal point
\usepackage{bm}% bold math
\usepackage{hyperref}% add hypertext capabilities
\usepackage{siunitx}
\usepackage{overpic}
\usepackage[noabbrev,nameinlink]{cleveref}

\usepackage[normalem]{ulem}
\usepackage{comment}
\usepackage{natbib}

\newcommand{\diff}[2]{\frac{\mathrm{d}#1}{\mathrm{d}#2}}
\newcommand{\pdiff}[2]{\frac{\partial#1}{\partial#2}}
\renewcommand{\vec}[1]{\bm{#1}}
\newcommand{\norm}[1]{\left\lvert #1\right\rvert}
\newcommand{\intd}[1]{~\mathrm{d}#1}
\newcommand{\cross}{\mathbin{\times}}
\newcommand{\x}{\vec{x}}
\newcommand{\X}{\vec{X}}
\newcommand{\y}{\vec{y}}
\newcommand{\h}{\vec{h}}
\newcommand{\hI}{\vec{h}_I}

\renewcommand{\tensor}[1]{\bm{#1}}
\newcommand{\ord}[1]{\operatorname{ord}\left(#1\right)}
\newcommand{\bigO}[1]{O\left(#1\right)}
\newcommand{\littleo}[1]{o\left(#1\right)}
\newcommand{\V}{\vec{V}}
\renewcommand{\u}{\vec{u}}
\newcommand{\m}{\vec{m}}
\newcommand{\R}{\vec{R}}
\newcommand{\et}{\vec{e}_t}
\newcommand{\en}{\vec{e}_n}
\newcommand{\eb}{\vec{e}_b}
\newcommand{\ex}{\vec{e}_x}
\newcommand{\ey}{\vec{e}_y}
\newcommand{\ez}{\vec{e}_z}
\newcommand{\ephi}{\vec{e}_{\phi}}
\newcommand{\er}{\vec{e}_r}
\newcommand{\vel}{\vec{a}}
\newcommand{\dVol}{\Delta_{\text{vol}}}

\newcommand{\eye}{\tensor{I}}
\newcommand{\f}{\vec{f}}
\newcommand{\T}{\vec{T}}

\newcommand{\Erotlet}{\mathcal{E}^{\text{rotlet}}}
\newcommand{\ERTT}{\mathcal{E}^{\text{RTT}}}

\let\oldxi\xi
\let\oldOmega\Omega
\let\oldalpha\alpha
\renewcommand{\xi}{\vec{\oldxi}}
\renewcommand{\Omega}{\vec{\oldOmega}}
\renewcommand{\alpha}{\vec{\oldalpha}}

\graphicspath{{./figs/}}

\begin{document}

\title{A hydrodynamic slender-body theory for local rotation at zero Reynolds number}

\author{Benjamin J. Walker}
\email{Corresponding author: bjw43@bath.ac.uk}
\affiliation{Department of Mathematical Sciences, University of Bath, Bath, BA2 7AY, UK}
\affiliation{Department of Mathematics, University College London, London, WC1H 0AY, UK}

\author{Kenta Ishimoto}
\email{ishimoto@kurims.kyoto-u.ac.jp}
\affiliation{Research Institute for Mathematical Sciences, Kyoto University, Kyoto, 606-8502, Japan}

\author{Eamonn A. Gaffney}
\email{gaffney@maths.ox.ac.uk}
\affiliation{Wolfson Centre for Mathematical Biology, Mathematical Institute, University of Oxford, Oxford, OX2 6GG, UK}

    
\date{\today}

\begin{abstract}
\input{sections/abstract}
\end{abstract}

\maketitle

% Main text.
\section{Introduction}
\label{sec: intro}
\input{sections/introduction}

\section{The rotating slender-body problem}
\label{sec: setup}
\input{sections/setup}

\section{Rotating slender bodies}
\label{sec: rotating}
\input{sections/rotating}

\section{Capturing translation}
\label{sec: translation}
\input{sections/translation}

\section{Numerical examples}
\label{sec: numerics}
\input{sections/numerics}

\section{Discussion}
\label{sec: discussion}
\input{sections/discussion}

\section*{Acknowledgements}
B.J.W.\ is supported by the UK Engineering and Physical Sciences Research
Council (EPSRC), grant EP/N509711/1, and the Royal Commission for the Exhibition of 1851. K.I.\ acknowledges JSPS-KAKENHI for Young Researchers (Grant No. 18K13456), JSPS-KAKENHI for Transformative Research Areas (Grant No. 21H05309) and JST, PRESTO, Japan (Grant No. JPMJPR1921).

\appendix
\input{sections/appendices}

\end{document}

%% file: sections/abstract.tex
%!TEX root=../main.tex

Slender objects are commonplace in microscale flow problems, from soft
deformable sensors to biological filaments such as flagella and cilia. Whilst
much research has focussed on the local translational motion of these slender
bodies, relatively little attention has been given to local rotation, even
though it can be the dominant component of motion. In this study, we explore
a classically motivated ansatz for the Stokes flow around a rotating slender
body via superposed rotlet singularities, which leads us to pose an
alternative ansatz that accounts for both translation and rotation. Through
an asymptotic analysis that is supported by numerical examples, we determine
the suitability of these flow ansatzes for capturing the fluid velocity at
the surface of a slender body, assuming local axisymmetry of the object but
allowing the cross-sectional radius to vary with arclength. In addition to
formally justifying the presented slender-body ansatzes, this analysis
reveals a markedly simple relation between the local angular velocity and the
torque exerted on the body, which we term resistive torque theory. Though
reminiscent of classical resistive force theories, this local relation is
found to be algebraically accurate in the slender-body aspect ratio, even
when translation is present, and is valid and required whenever local
rotation contributes to the surface velocity at leading asymptotic order.

%% file: sections/introduction.tex
%!TEX root=../main.tex

Capturing the fluid flow around a slender object at low Reynolds number has
been a goal of countless research efforts, motivating the development of a
range of theoretical approaches over the last seventy years. Perhaps best
known amongst these is the resistive force theory introduced by
\citet{Hancock1953,Gray1955} in the 1950s, which established a simple, local
approximation to the relationship between the velocity of a slender body and
the force that it exerts on the surrounding fluid. Since these seminal works,
resistive force theory has been iterated upon and refined into the broad class
of slender-body theories, which exploit the asymptotic slenderness of an
object to approximate and simplify the coupling between geometry and flow.
This class of theories includes refinements to resistive force theory in
complex environments \citep{Brenner1962,Katz1975}, as well as the integral
equations of \citet{wu1976,Johnson1977,Johnson1980,Keller1976,Higdon1979}.
More recently, the emergence of regularised Green's functions of Stokes flow
\citep{Cortez2001} has prompted the development of regularised slender-body
theories \citep{Walker2020b,Gillies2009,Cortez2012,Olson2013}, which are
typically simpler to implement in practice than singular theories and can even
afford additional flexibility \citep{Walker2020b,Gillies2009}, though at the
expense of a small regularisation error. Together, these slender-body theories
have been utilised in the study of a wide range of biological and biophysical
settings, such as in popular application to microscale motility and to the
dynamics of synthetic sensors
\citep{Guglielmini2012,Smith2009,Gillies2009,Roper2006,Olson2015}.

Broadly speaking, these slender-body theories seek to answer the following
question: given a slender object that is translating in a fluid, what are the
forces that are being exerted on the object as it moves relative to the
fluid? This problem can be easily motivated by the aforementioned examples,
with microscale swimming often involving significant side-to-side undulations
of slender bodies, such as flagella or cilia, in order to achieve locomotion.
Indeed, for slender objects that are not rapidly rotating, a statement that
we will make precise in \cref{sec: setup}, the dominant contribution to the
fluid flow at the surface of the body is from local translation, with the
slenderness of the object reducing the magnitude of any rotational effects.
However, if a slender object rotates sufficiently quickly, then local
rotation can have a greater effect on the fluid than translation, which is
readily seen to be the case for slender bodies that are not translating at
all. Though the examples mentioned above are not often associated with rapid
rotation, circumstances where rotation is significant arise in similar
contexts. For instance, the twisting of slender bodies has been explored in
detail in the context of filament instabilities, termed \emph
{twirling} and \emph{whirling} in the study \citet{Wolgemuth2000}, motivated
in part by the small-scale instabilities of DNA \citep{Liu1987}, though it is
unclear if the scales associated with DNA are always compatible with the
continuum fluid assumption. Further applications include ciliary
beating \citep{Holwill1979} and exploring the shapes adopted by polymorphic
bacterial flagella \citep{Macnab1977}. More recently, with the advent of
efficient elastohydrodynamic simulation methods \citep
{Walker2020a,Schoeller2019}, a need has arisen for accurate quantification of
the hydrodynamic torques that act on slender bodies in flow. For instance, in
a study of twisting and buckling, one can imagine imposing an arbitrarily
large axial twist on an elastic filament in silico. The subsequent relaxation
dynamics towards an untwisted equilibrium will then typically lead to rapid
rotation of the slender body, with accurate study thereby warranting careful
quantification of the fluid-imparted resistance to local rotation. Hence, the
contribution of rotational motion cannot be neglected in general, motivating
the development of a practical slender-body theory that is capable
of accurately resolving rotational effects.

However, to the best of our knowledge and in spite of the significant
attention given to translational theories, studies analysing, justifying and
validating a rotational slender-body theory have been absent until a very
recent study by \citet{Maxian2022}. However, the framework resulting from this
initial study suffers from an undetermined parameter and, additionally, the
restriction that the weighting of the Stokeslet in the single layer boundary
integral formulation is equal to the surface traction, which necessarily
assumes that the rotating body is constituted by a Newtonian fluid of the same
viscosity as the surrounding medium. Such a restriction is inappropriate for
cellular bioswimmers, for example due to the poroelasticity of eukaryotic
cytoplasm \citep{Moeendarbary2013}.

Thus, as the primary aim of this study, we will seek to pose and present a
detailed analysis of a slender-body theory that expressly accounts for
rotational motion whilst imposing relatively weak restrictions, focussing in
particular on regimes where angular velocity is sufficiently fast to
contribute to the motion of the object at leading asymptotic order. In doing
so, we will expand upon the rotation-capturing studies of \citet{Keller1976}
and \citet{Koens2018b} by exploring a practical theory that we will
demonstrate, both analytically and numerically, to be asymptotically accurate
and broadly applicable.

Our initial slender-body ansatz, which we will introduce in \cref {sec:
rotating}, will be strongly linked to the classical study of \citet
{Chwang1974}, which identified exact solutions for the flow around a range of
axisymmetric bodies that were not necessarily slender. Here, we will look to
relax this assumption of global axisymmetry, instead assuming slenderness and
a notion of local axisymmetry, which we introduce formally later. Our analysis
of this ansatz will motivate us to combine it with an existing, translational
slender-body theory, which we will demonstrate is capable of capturing both
the forces and torques acting on the body with asymptotic accuracy. Distinct
in complexity from the general theory of \citet {Koens2018b}, this combined
ansatz will resemble those employed in an ad hoc manner in recent
computational works \citep
{Flores2005,Nazockdast2017,Ishimoto2018a,Huang2019,Carichino2019}, though here
we will seek to formally justify its suitability and its accuracy, an
assessment that we believe to be absent from previous studies. Thus, a further
aim of this study will be to ascertain, to some degree, the suitability of a
combined ansatz for the study of the general motion of locally axisymmetric
slender bodies, incorporating both translation and rotation.

Hence, in this study, we will begin by precisely formulating the slender-body
problem for locally axisymmetric objects in flow. Initially restricting
ourselves to consideration of bodies that are undergoing local rotation only,
without accompanying translation, we will asymptotically determine the
suitability of this ansatz for accurately quantifying the torque exerted on
the object by the fluid. Following further analysis of a simple numerical
scheme suggested by this ansatz, we will move to consider motion that also
includes translation, carefully justifying the coupling of our
rotation-capturing ansatz with a traditional, translational slender-body
theory. We will explore this theory asymptotically, culminating in a simple,
local theory that is reminiscent of classical resistive force theories but
with surprising asymptotic accuracy, and validate our theoretical results
through numerical exploration.

%% file: sections/setup.tex
%!TEX root=../main.tex

\subsection{Geometry and governing equations}
Throughout, we will adopt the notation and set-up of \citet{Walker2020b},
which we briefly summarise in dimensionless form and illustrate in \cref{fig:
setup}. In this work, we will consider a three-dimensional, inextensible,
unshearable slender body in a Newtonian fluid, with the fluid domain being
taken to be the exterior of the slender body. In an inertial reference frame,
the velocity of this fluid $\u(\vec{x})$ at a point $\vec{x}$ is governed by
the familiar dimensionless Stokes equations
\begin{equation}
    \nabla \cdot \u = 0\,, \qquad \vec{0} = -\nabla p + \nabla^2 \u\,,
\end{equation}
where $p(\x)$ is the pressure. We will assume that the slender body is locally
axisymmetric, such that its cross-section at any point is circular and lies in
a plane transverse to the centreline \citep{Antman2005}. With this assumption,
the geometry of the body is entirely defined by its centreline $\xi(s)$ and
cross-sectional radius $\hat{\eta}(s)$, where $s\in[-1,1]$ is a material
arclength parameter. Here, $\hat{\eta}(s)$ is assumed to be non-negative,
vanishing only at $s=\pm 1$, in terms of which we can capture the slenderness
of the body via
\begin{equation}\label{eq:epsilon}
    \epsilon = \max_{s\in[-1,1]}\hat{\eta}(s)\,,
\end{equation}
with slenderness defined as the regime $\epsilon\ll1$, assumed throughout.
This prompts the definition $\eta(s) \coloneqq \hat{\eta}(s)/\epsilon$, so
that $\eta(s)$ is sharply bounded above by unity. In what follows, we will refer to this normalised function as the radius function.

\begin{figure}
    \centering
    \begin{overpic}[width=0.9\textwidth]{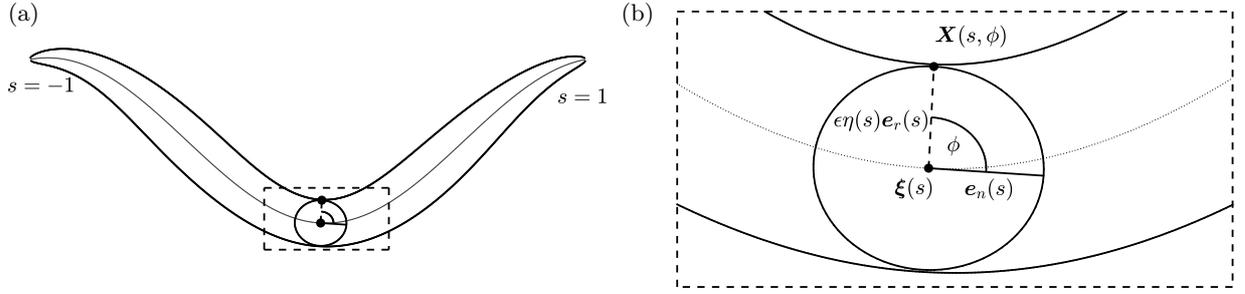}
    \put(0,22){(a)}
    \put(50,22){(b)}
    \end{overpic}
    \caption{A slender body with circular cross-section of
    arclength-dependent radius $\epsilon\eta(s)$, centreline $\xi(s)$,
    and cross-sectional angle $\phi$. Points on the surface of the body are
    parameterised as $\vec{X}(s,\phi)$, where $s\in[-1,1]$ is an arclength
    parameter. Figure adapted from \citet{Walker2020b} with permission.}
    \label{fig: setup}
\end{figure}

The centreline and cross-sections of the body may be further described with
the aid of the orthonormal triad $\{\et,\en,\eb\}$, comprised of
arclength-dependent unit tangent, normal, and binormal vectors to $\xi$. More
explicitly, these vectors satisfy the standard Frenet-Serret relations
\begin{equation}\label{eq:FrenetSerret}
    \et(s) = \pdiff{\xi}{s}, \qquad \frac{\partial \et}{\partial s} = \kappa(s) \en(s), \qquad
    \eb(s) = \et(s) \cross \en(s)\,,
\end{equation}
with $\kappa(s)$ being the centreline curvature. This local basis allows us to
define the radial unit vector $\er$, which lies in the transverse
cross-section of the body, in terms of the cross-sectional angle $\phi$ as
\begin{equation}\label{eq: er definition}
    \er(s,\phi) \coloneqq \en(s) \cos \phi + \eb(s) \sin \phi\,,
\end{equation}
with $\phi$ illustrated in \cref{fig: setup}b. With this notation, points
on the surface of the slender body may be succinctly parameterised as
\begin{equation}\label{eq:surfacepoint}
    \vec{X}(s,\phi) = \xi(s) + \epsilon\eta(s) \er(s,\phi)\,.
\end{equation}

\subsection{Boundary conditions}
As is standard in slender-body applications, we will seek to apply the
hydrodynamic no-slip condition on the surface of the slender body. For fluid
velocity $\u(\vec{X}(s,\phi))$ at a surface point parameterised by
arclength $s$ and angle $\phi$, we accordingly impose
\begin{equation}
    \u(\vec{X}(s,\phi)) = \diff{\vec{X}(s,\phi)}{t}\,,
\end{equation}
where $t$ denotes dimensionless time. The velocity of the surface, which we will assume conserves the volume of the slender body, may be
decomposed as
\begin{equation}
    \diff{\vec{X}(s,\phi)}{t} = \V(s) + \epsilon\eta(s)\vec{\Omega}(s) \cross \er(s,\phi)\,.
\end{equation}
Here, $\V(s)$ is the linear or translational velocity and $\vec{\Omega}(s)$
is the rate of local rotation about the centre of the cross section. Of note, this form includes, but is not limited to, rigid-body motion of the slender object. For completeness, the
general no-slip condition may be stated as
\begin{equation}\label{eq: general BC}
    \u(\vec{X}(s,\phi)) = \V(s) + \epsilon\eta(s)\vec{\Omega}(s)\cross \er(s,\phi)\,.
\end{equation}

%% file: sections/rotating.tex
%!TEX root=../main.tex

In this first exploration, we will adopt a simple, classically motivated
ansatz and explore the efficacy of using it to satisfy the velocity boundary
condition on slender bodies that are undergoing only rotation, i.e. $\V(s) =
\vec{0}$.

\subsection{A rotlet ansatz}
Consider the following ansatz for the flow around a rotating slender body at a point $\x$:
\begin{equation}\label{eq: pure rotlet ansatz}
    \u_R(\x) = \int_{-e}^e \m(s') \cross \underbrace{\R(\x,\xi(s'))}_{\text{rotlet}}\intd{s'}\,,
\end{equation}
where $e\coloneqq(1-\epsilon^2)^{1/2}$ defines a notional eccentricity and the
\emph{rotlet} $\R$ is given by
\begin{equation}
    \R(\x,\y) \coloneqq \frac{\x - \y}{\norm{\x - \y}^3}\,.
\end{equation}
With these definitions, we have that $8\pi\m(s)$ is the torque density
imparted on the fluid by the body, noting that we have non-dimensionalised the
viscosity to unity, in line with Equation 9 of \citet{Chwang1975}. This rotlet
is a fundamental singularity solution of Stokes equations, corresponding to
the flow induced by point torque \cite {Kim2005}. Hence, this ansatz
represents the flow due to a line distribution of point torques, motivated by
the classical study of \citet {Chwang1974}, which employed a similar ansatz to
study the rotation of axisymmetric bodies. Here, we have taken the limits of
integration to be $\pm e$ for later convenience, though note that they can
readily be substituted for $\pm1$, representing only a minimal difference as
$e\sim 1- \epsilon^2/2 +
\bigO{\epsilon^4}$ as $\epsilon\to0$.

We will look to determine the suitability of this torque-only ansatz for satisfying a purely rotational boundary condition, so that we will impose
\begin{equation}
    \u_R(\X(s,\phi)) = \epsilon\eta(s)\Omega(s)\cross\er(s,\phi)\,,
\end{equation}
having set $\V(s)=\vec{0}$ in \cref{eq: general BC}. Of note, this boundary
condition and ansatz, up to the limits of integration, are exactly of the form
studied by \citet{Chwang1974} if we restrict our consideration to (1) straight
slender bodies and (2) purely axial angular velocities, i.e.
$\Omega(s)\parallel\et(s)$. Here, we will seek to retain generality in both
the shape of the slender body and the considered angular velocity, with our
assumptions of body slenderness and circular cross-sections in the transverse
plane replacing \citeauthor{Chwang1974}'s assumption of axisymmetry.

\subsection{Resistive torque theory}
\label{sec: resistive torque theory}
In order to evaluate the suitability of the ansatz of \cref{eq: pure rotlet ansatz}, we will pursue an asymptotic analysis in the slenderness parameter $\epsilon\ll1$. Requiring that the ansatz of \cref{eq: pure rotlet ansatz} satisfies the rotational boundary condition on the surface of the slender body yields the relation
\begin{equation}\label{eq: rotational boundary condition ansatz}
    \epsilon\eta(s)\Omega(s)\cross\er(s,\phi) = \int_{-e}^e \underbrace{\m(s') \cross \frac{\X(s,\phi) - \xi(s')}{\norm{\X(s,\phi) - \xi(s')}^3}}_{\h(s,s',\phi)} \intd{s'}\,,
\end{equation}
which must hold for all $s\in[-1,1]$ and $\phi\in[0,2\pi)$. If we denote the
integrand by $\h(s,s',\phi)$, we can make progress in evaluating the integral
by noting that $\norm{\h}=\bigO{\norm{\m}}$ when $s-s'=\ord{1}$\footnote{That
is, $s-s'$ is both $\bigO{1}$ and not $\littleo{1}$.}, and
$\norm{\h}=\bigO{\norm{\m}/\epsilon^2}$ when $s-s'=\bigO{\epsilon}$. Hence,
the contributions of these regions to the integral are $\bigO{\norm{\m}}$ and
$\bigO{\norm{\m}/\epsilon}$, respectively. The difference in the size of the
integrand between these two regions leads us to adopt the terminology of
matched asymptotics, with $s-s'=\bigO{\epsilon}$ defining the so-called inner
region. In this case, asymptotic approximation of the rotlet integral is
particularly straightforward, with only the inner region contributing to the
integral at leading order in $\epsilon$. In this region, we may expand $\h$ in
terms of an inner variable $\sigma\coloneqq (s-s')/\epsilon = \bigO{1}$ as
\begin{equation}\label{eq: leading order expansion}
    \h(s,s',\phi) \sim \m(s)\cross\frac{\left[\epsilon\eta(s)\er(s,\phi) - \epsilon\sigma\et(s)\right]}{\left[\epsilon^2\sigma^2 + \epsilon^2\eta^2(s)\right]^{3/2}}\left(1 + \bigO{\epsilon}\right)\,,
\end{equation}
denoting this leading-order inner expansion by $\hI(s,s',\phi)$ and making use
of the local approximation $\xi(s) - \xi(s') \sim \epsilon\sigma\et(s) +
\bigO{\epsilon^2}$. This expansion of $\h$ is asymptotic if both
$\mathrm{d}\m/\mathrm{d}s$ and $\kappa$ are $\bigO{1}$, which each appear in
terms that are $\bigO{\epsilon^2}$. In terms of the integration variable $s'$,
the leading-order inner expansion of \cref{eq: leading order expansion} is
\begin{equation}
    \hI(s,s',\phi) \coloneqq \m(s)\cross \frac{\left[\epsilon\eta(s)\er(s,\phi) - (s-s')\et(s)\right]}{\left[(s-s')^2 + \epsilon^2\eta^2(s)\right]^{3/2}}\,,
\end{equation}
which decays like $1/(s-s')^2$ as $s'$ moves away from the inner region. In particular, this inner expansion is only $\bigO{1}$ outside of the inner region, so that we can in fact write
\begin{equation}\label{eq: int of f is int of fI}
    \int_{-e}^e\h(s,s',\phi)\intd{s'} \sim \int_{-e}^e\hI(s,s',\phi)\intd{s'}
\end{equation}
with error scaling with $\epsilon$, as the integral of $\hI$ outside the inner region is subleading. Thus, in order to evaluate the leading-order contribution of our ansatz, we need only evaluate the integral of $\hI$. This integral decomposes into
\begin{equation}\label{eq: hI decomposed}
    \int_{-e}^e\hI(s,s',\phi)\intd{s'} = \m(s)\cross\left[\epsilon\eta(s)I_1(s)\er(s,\phi) - I_2(s)\et(s)\right]\,,
\end{equation}
where
\begin{subequations}
\begin{align}
    I_1(s) &\coloneqq \int_{-e}^e \frac{1}{\left[(s-s')^2 + \epsilon^2\eta^2(s)\right]^{3/2}}\intd{s'}\,,\\
    I_2(s) &\coloneqq \int_{-e}^e \frac{s-s'}{\left[(s-s')^2 + \epsilon^2\eta^2(s)\right]^{3/2}}\intd{s'}\,.
\end{align}
\end{subequations}
For completeness, the required no-slip boundary condition reads
\begin{equation}\label{eq: leading order rotational BC}
    \epsilon\eta(s)\Omega(s)\cross\er(s,\phi) = \m(s)\cross\left[\epsilon\eta(s)I_1(s)\er(s,\phi) - I_2(s)\et(s)\right]
\end{equation}
to leading order. Immediately, \cref{eq: leading order rotational BC} presents cause for concern: the term containing $I_2(s)$ is independent of $\phi$, so represents a \emph{translational} velocity generated by our ansatz, which is not compatible with the angular velocity boundary condition. However, for now, it is instructive to delay the consideration of this issue and proceed to try to relate $\m(s)$ to $\Omega(s)$ more simply.

In particular, we recall that the boundary condition must hold for all $s$ and $\phi$, so that it must simultaneously hold for any given $\phi$ and $\phi+\pi$ at a fixed $s$. Therefore, noting that $\er(s,\phi+\pi)=-\er(s,\phi)$ from \cref{eq: er definition}, we have the simultaneous equations
\begin{subequations}
\begin{align}
    \epsilon\eta(s)\Omega(s)\cross\er(s,\phi) &= \hphantom{-}\m(s)\cross\left[\epsilon\eta(s)I_1(s)\er(s,\phi) - I_2(s)\et(s)\right]\,,\\
    -\epsilon\eta(s)\Omega(s)\cross\er(s,\phi) &= -\m(s)\cross\left[\epsilon\eta(s)I_1(s)\er(s,\phi) + I_2(s)\et(s)\right]\,.
\end{align}
\end{subequations}
Notably, as the induced translational velocity is independent of $\phi$, it is unaltered by this change in cross-sectional angle. Hence, this contribution cancels upon taking the difference between the two equations, yielding the relation
\begin{equation}\label{eq: almost rotating rrt}
    \Omega(s)\cross\er(s,\phi) = I_1(s)\m(s)\cross\er(s,\phi)\,,
\end{equation}
valid to leading order unless $\eta(s)=0$, which can only occur at the ends of the slender body. As $\phi$ is arbitrary in \cref{eq: almost rotating rrt} and $\er(s,\phi)$ and $\er(s,\phi+\pi/2)$ are linearly independent, this equality between vector products allows us to conclude that 
\begin{equation}
    \Omega(s) = I_1(s)\m(s)\,,
\end{equation}
so that $\Omega(s)$ is proportional to $\m(s)$ with scale factor $I_1(s)$. This is a leading-order \emph{local} relation between the rotation of a general slender body and the hydrodynamic torque that acts on it, which we term \emph{resistive torque theory} (RTT).

Further, we can explicitly evaluate $I_1(s)$ as 
\begin{equation}\label{eq: I1 exact}
    I_1(s) = \left[\frac{s'-s}{\epsilon^2\eta^2(s)\left[(s-s')^2 + \epsilon^2\eta^2(s)\right]^{1/2}}\right]_{-e}^e\,.
\end{equation}
This corresponds exactly to the result of \citet[sec. 4]{Chwang1974} for
straight, constant-torque bodies, highlighting that our leading-order
asymptotic result is analogous to treating the slender body as locally
straight with a constant torque density. 

\begin{figure}
    \centering
    \includegraphics[width = 0.6\textwidth]{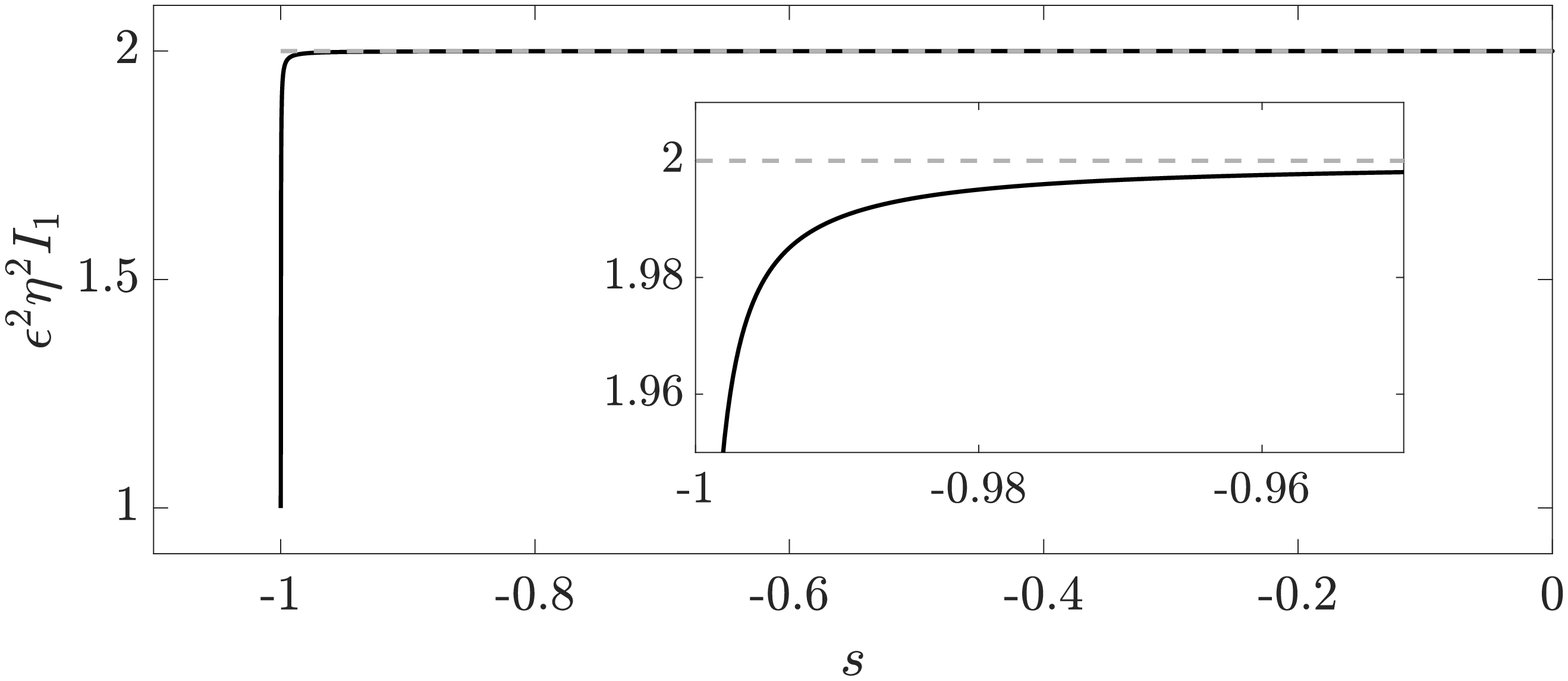}
    \caption{Approximation of $I_1(s)$. We illustrate $I_1(s)$ and a leading-order approximation to it, each normalised by $1/\epsilon^2\eta^2(s)$. The exact value of $I_1(s)$ is shown as a solid black curve, with the approximation of \cref{eq: I1 approx in middle} shown as a grey dashed line. We observe excellent agreement between the approximation of \cref{eq: I1 approx in middle} and the true value of $I_1(s)$ for the vast majority of the slender body, with the discrepancy towards the ends highlighted in the inset. Here, we have taken $\eta(s) = (1 - s^2)^{1/2}$ and $\epsilon=10^{-2}$, corresponding to a slender prolate ellipsoid, though we note that the qualitative features of this figure are not sensitive to this choice.}
    \label{fig: I1}
\end{figure}

In our slender limit, we can further simplify the expression for $I_1(s)$. In
particular, we note that we can write
\begin{equation}
    \epsilon^2\eta^2(s)I_1(s) = F(e-s) + F(e+s)\,, \quad F(t) \coloneqq \frac{t}{\left(t^2 + \epsilon^2\eta^2(s)\right)^{1/2}}\,.
\end{equation}
For $s\in[-1,1]$ that are not too close to the ends of the body or,
more precisely, those $s$ such that $\epsilon^2\eta^2(s)(e+s)^{-2} =
\bigO{\epsilon}$ and $\epsilon^2\eta^2(s)(e-s)^{-2} = \bigO{\epsilon}$, we can asymptotically expand $F(e-s)$ and $F(e+s)$, yielding $F(e-s) \sim F(e+s) = 1 + \bigO{\epsilon}$, so that 
\begin{equation}\label{eq: I1 approx in middle}
    I_1(s) \sim \frac{2}{\epsilon^2\eta^2(s)}\left(1 + \bigO{\epsilon}\right)\,.
\end{equation}
For $s$ approaching $-e$, we write $s = -e + \delta$ for some $\delta$ such that
\begin{equation}
    0 \leq \delta \ll \epsilon\eta \ll 1\,,
\end{equation}
noting that $\eta(-e) > 0$ by assumption. Here, we have $F(e-s) = F(2e-\delta) = 1 + \bigO{\epsilon^2\eta^2(s)}$, completely analogous to the case when $s$ was away from the ends of the body. However, expanding $F(e+s)$ now yields
\begin{equation}
    F(e+s) = F(\delta) = \frac{\delta}{\epsilon\eta(s)}\left(1 + \bigO{\frac{\delta^2}{\epsilon^2\eta^2(s)}} \right) = \frac{s+e}{\epsilon\eta(s)}\left(1 + \littleo{1}\right)\,.
\end{equation}
Hence, with $\delta = \bigO{\epsilon^{3/2}\eta}$, we have
\begin{equation}\label{eq: I1 approx near ends}
    I_1(s)\sim \frac{1 + \frac{s+e}{\epsilon\eta(s)}}{\epsilon^2\eta^2(s)}\left(1 + \bigO{\epsilon}\right)
\end{equation}
as $s$ approaches $-e$, with an analogous result holding as $s$ approaches $e$. The leading-order approximation to $I_1(s)$ of \cref{eq: I1 approx in middle} is illustrated in
\cref{fig: I1}, highlighting excellent agreement with the exact expression of
\cref{eq: I1 exact} away from the ends of the body. Hence, for the vast
majority of the slender object, where \cref{eq: I1 approx in middle} holds, we
have the simple leading-order resistive torque theory relation
\begin{equation}\label{eq: rotating rtt}
    \frac{\epsilon^2\eta^2(s)}{2}\Omega(s) = \m(s)\,.
\end{equation}
This resistive torque theory appears to hold for all components of $\Omega(s)$
and $\m(s)$ with the same coefficient of proportionality, in stark contrast to
the anisotropic resistive force theory coefficients associated with the
translation of slender objects in a viscous fluid. Further, if we were to restrict the rotation and the associated torque to be only in the axial direction, this is precisely in line with the relation found by \citet{Keller1976}. It is also useful to note that this result gives ready access to an analogue of a rotational drag coefficient for axial rotations of a straight slender body, with the relationship between an imposed, arclength-independent, axial rotation $\Omega$ and the total torque on the body being readily accessible through integration of \cref{eq: rotating rtt} in space. Explicitly, in terms of a total applied torque $\T$, we have the leading-order relation
\begin{equation}
    \T = 8\pi \int_{-e}^{e}\m(s)\intd{s} = 4\pi\epsilon^2\Omega\int_{-e}^{e}\eta^2(s)\intd{s}\,,
\end{equation}
recalling that we are in a dimensionless regime with dimensionless viscosity
of unity.

\subsection{Preventing unwanted translation}
However, whilst the resistive torque theory of \cref{eq: rotating rtt} is somewhat beguiling, we should no longer ignore the undesirable translational velocity associated with our rotlet ansatz, which we recall is generated by the product $\m(s)\cross I_2(s)\et(s)$ in \cref{eq: leading order rotational BC}. In order for our ansatz (and our derived resistive torque theory) to be suitable for satisfying the rotational boundary condition on our slender body, we need this term to vanish for each $s$, so that we must have either $I_2(s)=0$ or $\m(s)\cross\et(s)=\vec{0}$. However, we can directly evaluate $I_2(s)$ as
\begin{equation}\label{eq: I2 exact}
    I_2(s) = \left[\frac{-1}{\left[(s-s')^2 + \epsilon^2\eta^2(s)\right]^{1/2}}\right]_{-e}^e\,,
\end{equation}
which vanishes only at $s=0$ and generally scales as $I_2(s)=\bigO{1/\epsilon\eta(s)}$. Thus, to remove the unwanted translational velocities in general, it must be the case that $\m(s)\cross\et(s)=\vec{0}$ whenever $s\neq0$, so that the torque $\m(s)$ is almost-everywhere parallel to the local tangent $\et(s)$. Recalling that our resistive torque theory of \cref{eq: rotating rtt} states that $\m(s)$ is parallel to $\Omega(s)$, this condition is equivalent (at leading order) to requiring that the rotation of the slender body is purely axial, so that $\Omega(s)$ is parallel to $\et(s)$. Hence, the pure-rotlet ansatz examined in this section, and the resistive torque theory that it generates, is justified for use only for slender bodies that are locally rotating about their centreline.

\subsection{Invertibility of the discretised rotlet ansatz}\label{sec: invertibility}
Given an angular velocity $\Omega(s)$, an approach for determining the unknown
torque density $\m(s)$ would be to discretise the centreline of the slender
body into $N$ segments of length $2e/N$, approximating $\m(s)$ as being
constant on each segment. Despite the simplicity of this approach, it is then
unclear how best to apply the angular velocity boundary condition, as directly
imposing the surface velocity at one randomly point on each of the $N$
segments can yield singular linear systems, which we reason arises due to the
presence of the vector product in the integral kernel of \cref{eq: pure rotlet
ansatz}. However, by following the reasoning that enabled us to invert
\cref{eq: almost rotating rrt}, we can apply the boundary condition in a way
that will provide sufficient information to solve for $\m(s)$ uniquely,
subject to a condition on $\epsilon$ and $N$.

In more detail, fixing an arclength $s=s_i$ on the $i$\textsuperscript{th}
segment, we consider the three projections
\begin{subequations}
\label{eq: rotlet bc projection}
\begin{align}
    \u_R(\X(s_i,0)) \cdot \et(s_i) &= \int_{-e}^e \m(s') \cross \R(\X(s_i,0), \xi(s')) \intd{s'} \cdot \et(s_i)\,,\\ 
    \u_R(\X(s_i,0)) \cdot \eb(s_i) &= \int_{-e}^e \m(s') \cross \R(\X(s_i,0), \xi(s')) \intd{s'} \cdot \eb(s_i)\,,\\ 
    \u_R(\X(s_i,\pi/2)) \cdot \et(s_i) &= \int_{-e}^e \m(s') \cross \R(\X(s_i,\pi/2), \xi(s')) \intd{s'} \cdot \et(s_i)\,,
\end{align}
\end{subequations}
of the full velocity boundary condition, which is being applied at the two
points parameterised by $(s_i,0)$ and $(s_i,\pi/2)$\footnote{Our choice of
$\phi=0$ here is for notational convenience only, with the argument readily
modified to consider general $\phi$.}. The motivation behind our choice of
these three projections becomes clear when considering the leading-order
approximations to the rotlet integrals, which, following our expansion of
\cref{eq: hI decomposed}, yields the leading-order form of \cref{eq: rotlet bc
projection} as
\begin{subequations}
\label{eq: leading order rotlet bc projection}
\begin{align}
    \u_R(\X(s_i,0)) \cdot \et(s_i) &= -\epsilon\eta(s_i)I_1(s_i)\eb(s_i)\cdot\m(s_i)\,,\\ 
    \u_R(\X(s_i,0)) \cdot \eb(s_i) &= \left[\epsilon\eta(s_i)I_1(s_i)\et(s_i) + I_2(s_i)\en(s_i)\right]\cdot\m(s_i)\,,\\ 
    \u_R(\X(s_i,\pi/2)) \cdot \et(s_i) &= -\epsilon\eta(s_i)I_1(s_i)\en(s_i)\cdot\m(s_i)\,,
\end{align}
\end{subequations}
appealing to the cyclic property of the scalar triple product and the
definition of $\er(s,\phi)$ from \cref{eq: er definition}. In particular, it
is clear that these local relations uniquely specify $\m(s_i)$ to leading
order, so that the corresponding $3\times3$ matrix representation is
invertible. Hence, the block-diagonal linear system formed by imposing these
leading-order relations for all $i=1,\ldots,N$ would itself be invertible,
constructed from invertible blocks. Strictly, this requires that $\epsilon$ be
smaller than $2e/N$, so that the leading-order contributions of the rotlet
integrals are associated with a single discrete $\m(s_i)$.

This result does not immediately imply the invertibility of \cref{eq: rotlet
bc projection}, as the linear system formed from discretising the integrals of
\cref{eq: rotlet bc projection} would have non-zero off-diagonal blocks.
However, fixing $i$ and assuming that $\epsilon\ll 2e/N$, each of the
off-diagonal blocks is strictly a factor of $\epsilon$ smaller than the
diagonal block (in any suitable matrix norm), which follows from \cref{eq:
leading order expansion} and the subsequent analysis. Hence, as there are
$N-1$ such blocks, the total magnitude of the off-diagonal blocks scales with
$\epsilon N$ relative to the diagonal block, which we recall is invertible.
Therefore, as this reasoning applies for each $i$, the resulting linear system
is \emph{block diagonally dominant}, following the definition of
\citet{Feingold1962}, if $\epsilon N < 1$. Thus, the block-matrix
generalisation of the Gerschgorin circle theorem of \citet{Feingold1962}
guarantees that the eigenvalues of the system are bounded away from zero, so
that the system is invertible. Therefore, up to $\bigO{1}$ constants, we in
fact have that the discretised rotlet ansazt, when implemented as described
above, yields invertible linear systems if $\epsilon N < 1$, establishing an
approximate upper bound for $N$ for guaranteed invertibility.

%% file: sections/translation.tex
%!TEX root=../main.tex

In the previous section, we saw that a flow ansatz comprised of only a
distribution of rotlets was able to satisfy an angular-velocity boundary
condition on the surface of a slender body, though the emergence of a
torque-induced translational velocity limited the applicability of the ansatz
to bodies that only rotate about their local tangent. In this section, we will
seek to remove this restriction by augmenting the rotlet ansatz with a
traditional slender-body theory ansatz that is capable of capturing
translational motion. 

However, the majority of slender-body theories are tailored to capturing
translational surface velocities that preserve the volume of the slender body,
invariably being a superposition of divergence-free singularities of Stokes
flow. Indeed, these theories are only capable of representing
volume-conserving flows \citep{Koens2018b}. Although the individual rotlet is
volume-conserving, and hence so is the rotlet induced translational flow from
the rotlet distribution of \cref{eq: pure rotlet ansatz}, this is not true for
the general arclength-dependent imposed velocity field, $\bm V(s)$ in  \cref
{eq: general BC}. Thus, we consider the constraints required for volume
conservation, at least to $\bigO {\epsilon}$, the level of asymptotic accuracy
to which we will be working.

\subsection{Volume conservation}
\label{sec: volume conservation}
To proceed, we consider an arclength dependent velocity field $\vel(s)$ and
its associated net volume flux, defined by 
\begin{equation}\label{eq: net volume flux def}
    \dVol (\vel) \coloneqq \iint \vel(s)\cdot\intd{\vec{S}} = \int_{-e}^e \int_0^{2\pi} \vel(s)\cdot \left(\pdiff{\X}{\phi}\cross\pdiff{\X}{s}\right)\intd{\phi}\intd{s} \,,
\end{equation}
where $\X(s,\phi)$ is as defined in \cref{eq:surfacepoint}, we have selected the outward-pointing surface normal, and we are integrating the volume flux over the entire surface of the slender body. In \cref{app: vol flux}, we show that
\begin{equation}\label{eq: dVol}
    \dVol(\vel) 
    = 2\pi\epsilon^2\left[  \bigO{\kappa\sup_{s}[\eta^2\norm{\vel}]}+\bigO{ \sup_{s}[\norm{\eta\eta'}\norm{\vel\cdot\bm e_t}]} \right],
\end{equation}
with $\eta' = \mathrm{d}\eta/\mathrm{d}s.$ 

Hence, if a translational velocity $\vel(s) = \V(s)=\bigO{1}$ is imposed
in \cref{eq: general BC}, the constraints $\kappa = \bigO{1}$ and
$\eta^2\eta'\V\cdot \et = \bigO{1}$ ensure volume conservation to $\bigO
{\epsilon}$, with a correction at $\bigO{\epsilon^2}$. In turn, these
constraints require that the curvature is not too large, that the radius
function does not vary too rapidly, and that the tangential velocity
component is not too large. Finally, note that with $\kappa=\bigO{1}$ once more and
taking 
\begin{equation}
    \vel (s) = \m(s)\cross I_2(s)\eta(s)\et(s)\,,
\end{equation}
so that $\vel \cdot \et = 0$ and with $\norm{\m} = O(\epsilon)$, as confirmed a posteriori below via \cref{eq: general full} and \cref{eq: general rtt}, we have $\norm{\vel}= \bigO{1}$ and
\begin{equation}
    \dVol(\m(s)\cross I_2(s)\eta(s)\et(s))  = \bigO{\epsilon^2}\,.
\end{equation} 
This also provides an explicit self-consistency check that the asymptotic
approximation to the translational flow velocity induced by the rotlet ansatz
with the above curvature constraint also satisfies volume conservation to $\bigO{\epsilon}$.

\subsection{A combined ansatz}\label{sec: combined ansatz}
As the induced translational velocity from the rotlet ansatz satisfies the
condition of volume conservation to leading order in $\epsilon$, we are able
to account for it in our boundary condition via the addition of a
translational slender-body theory. There are a wide range of candidate
theories, including many of those touched upon in \cref{sec: intro}, though we
will opt to use the recent theory of \citet{Walker2020b} as it both shares the
notation used so far in this manuscript and generates volume-conserving
translational surface velocities with algebraic asymptotic errors. Their
translational ansatz is
\begin{equation}
    \u_T(\x) \coloneqq \int_{-e}^e \left[\tensor{S}^{\chi(s')}(\x,\xi(s')) - \frac{1-e^2}{2e^2}(e^2 - s'^2)\tensor{D}^{\chi(s')}(\x,\xi(s'))\right]\f(s')\intd{s'}\,,
\end{equation}
whose limits of integration motivate our earlier choice of limits in \cref{eq:
pure rotlet ansatz}. Here, $8\pi\f(s)$ is the force density applied to the
fluid by the body, analogous to the torque density of the rotlet ansatz. The
tensors $\tensor{S}^{\chi(s')}$ and $\tensor{D}^{\chi(s')}$ correspond to
regularised Stokeslets and potential dipoles of Stokes flow
\citep{Cortez2001}, respectively, with a shared regularisation parameter
$\chi(s)\coloneqq \epsilon^2[1-s^2-\eta^2(s)]$. For completeness, these are of
the form
\begin{subequations}
\begin{align}
    \tensor{S}^\chi(\x,\y) &= \frac{(\norm{\x-\y}^2 + 2\chi)\eye}{(\norm{\x-\y}^2 + \chi)^{3/2}} + \frac{(\x-\y)\otimes(\x-\y)}{(\norm{\x-\y}^2 + \chi)^{3/2}}\,,\\
    \tensor{D}^\chi(\x,\y) &= -\frac{(\norm{\x-\y}^2 - 2\chi)\eye}{(\norm{\x-\y}^2 + \chi)^{5/2}} + \frac{3(\x-\y)\otimes(\x-\y)}{(\norm{\x-\y}^2 + \chi)^{5/2}}\,,
\end{align}
\end{subequations}
where $\eye$ is the identity tensor. We refer the interested reader to the
original publication of \citet{Walker2020b} for full details of their
regularised slender-body theory. Here, we only note a key result of
\citeauthor{Walker2020b}'s study: subject to $\kappa$ and $\eta$ not varying
too rapidly\footnote{The restriction on $\eta$ imposed by \citet{Walker2020b}
was later refined by \citet{Walker2021}, who noted that $\eta^2$ being
Lipschitz continuous with an $\bigO{1}$ constant was sufficient.}, the
leading-order flow that results from their ansatz at the surface of the
slender body is independent of $\phi$, so that it is purely translational. In
symbols, this can be summarised as
\begin{equation}\label{eq: uT expansion}
 \u_T(\X(s,\phi)) \sim \u_{T0}(s)\left(1 + \bigO{\epsilon}\right)
\end{equation}
for leading-order velocity $\u_{T0}$, where the correction term depends on both $s$ and $\phi$ in general.

These properties, along with our analysis of the pure rotlet ansatz, motivate the following combined ansatz, suitable for a slender body with the general rigid-body velocity of \cref{eq: general BC}:
\begin{multline}\label{eq: combined ansatz}
    \u_C(\x) \coloneqq \u_T(\x) + \u_R(\x) \\ = \int_{-e}^e \left[\tensor{S}^{\chi(s')}(\x,\xi(s')) - \frac{1-e^2}{2e^2}(e^2 - s'^2)\tensor{D}^{\chi(s')}(\x,\xi(s'))\right]\f(s')\intd{s'} \\ 
    + \int_{-e}^e \m(s') \cross \R(\x,\xi(s'))\intd{s'}\,. 
\end{multline}
Evaluated at the surface of the slender body and imposing the no-slip
condition of \cref{eq: general BC}, the boundary condition associated with
this ansatz reads
\begin{multline}\label{eq: combined ansatz full bc}
    \V(s) + \epsilon\eta(s)\Omega(s)\cross\er(s,\phi) =\\ \int_{-e}^e \left[\tensor{S}^{\chi(s')}(\X(s,\phi),\xi(s')) - \frac{1-e^2}{2e^2}(e^2 - s'^2)\tensor{D}^{\chi(s')}(\X(s,\phi),\xi(s'))\right]\f(s')\intd{s'} \\ 
    + \int_{-e}^e \m(s') \cross \R(\X(s,\phi),\xi(s'))\intd{s'}\,,
\end{multline}
with a typical slender-body problem then being to find the unknown force and torque densities given the translational and angular velocities. Owing to the established accuracy of the constituent translational and rotational ansatzes, the asymptotic accuracy of the combined ansatz is $\bigO{\epsilon\norm{\V} + \epsilon^2\eta\norm{\vec{\Omega}}}$, so that the combined error relative to the surface velocity is $\bigO{\epsilon}$.

\subsection{Enforcing no-slip in practice}
\label{sec: enforcing no-slip}
If one were to seek to solve the integral equation of \cref{eq: combined
ansatz full bc} numerically, a quadrature rule might be employed to compute
the integrals of the Stokeslet, potential dipole, and rotlet kernels. However,
the combined ansatz and the associated integral equation can be significantly
simplified using the results and principles of our analysis of the pure rotlet
ansatz in \cref{sec: rotating}, reducing the associated numerical cost.
Specifically, we can bypass numerical evaluation of the rotlet integral
entirely by applying the asymptotic result of \cref {sec: rotating}, removing
the need for quadrature for these integrals, though the integral involving the
Stokeslet and potential dipole in \cref{eq: combined ansatz full bc} is always
treated numerically in this study. In particular, retaining the leading-order
rotlet terms, with relative errors of $\bigO{\epsilon}$, yields
\begin{equation}\label{eq: partially reduced constraint}
    \V(s) + \epsilon\eta(s)\Omega(s)\cross\er(s,\phi) =\\
    \u_T(\X(s,\phi)) + \epsilon\eta(s)I_1(s)\m(s)\cross\er(s,\phi) - I_2(s)\m(s)\cross\et(s)\,,
\end{equation}
recalling that $I_1(s)$ and $I_2(s)$ are known in closed form from \cref{eq:
I1 exact,eq: I2 exact}. With this approximate form, evaluating the
contribution of the torque terms is akin to evaluating a resistive torque
theory in terms of numerical cost.

A further theoretical simplification can be made by appropriately selecting
the surface points at which to impose the no-slip condition. Inspired by our
resistive torque theory result of \cref{sec: rotating}, we are motivated to
apply the boundary condition at the pairs of points $(s,\phi)$ and
$(s,\phi+\pi)$, with any $\phi$-independent terms only needing to be computed
once per pair. Imposing the condition of \cref{eq: partially reduced
constraint} at these points is equivalent to imposing \cref{eq: partially
reduced constraint} at $(s,\phi)$ along with the modified constraint
\begin{equation}\label{eq: difference boundary condition}
    2\epsilon\eta(s)\Omega(s)\cross\er(s,\phi) =\\
    \left[\u_T(\X(s,\phi))-\u_T(\X(s,\phi+\pi))\right] + 2\epsilon\eta(s)I_1(s)\m(s)\cross\er(s,\phi)
\end{equation}
at $(s,\phi)$. This equivalence arises from noting that the latter condition is the result of taking the difference between \cref{eq: partially reduced constraint} evaluated at both $(s,\phi)$ and $(s,\phi+\pi)$. \Cref{eq: difference boundary condition} exploits the $\phi$-independence of the imposed translational velocity and the rotlet-induced translation to eliminate these terms from the boundary condition, mimicking the reasoning employed in \cref{sec: rotating} to generate our resistive torque theory.

\subsection{Generalised resistive torque theory}\label{sec: generalised resistive torque theory}
\Cref{eq: difference boundary condition} closely resembles an intermediate step in deriving the resistive torque theory of \cref{sec: resistive torque theory}, akin to \cref{eq: almost rotating rrt}. The additional term in \cref{eq: difference boundary condition}, $\u_T(\X(s,\phi))-\u_T(\X(s,\phi+\pi))$, represents the effects of the sub-leading couple generated by the forces distributed along the body centreline, contributing to the angular velocity along with the local torque through \cref{eq: difference boundary condition}. Hence, this equation captures the interactions between angular velocity, directly applied torques, and the sub-leading force-induced couple.

In pursuit of further simplification of this relation, we can estimate the magnitudes of the force- and torque-induced terms in \cref{eq: difference boundary condition}. Recalling the expansion of \cref{eq: uT expansion} established by \citet{Walker2020b} for the translational slender-body theory ansatz, the force-induced terms scale as
\begin{equation}
    \u_T(\X(s,\phi))-\u_T(\X(s,\phi+\pi)) = \bigO{\epsilon \norm{\u_{T0}}}\,,
\end{equation}
though this is a crude upper bound as gross cancellation may occur between $\u_T(\X(s,\phi))$ and $\u_T(\X(s,\phi+\pi))$. Furthermore, $\norm{\u_{T0}}$ is bounded by both the prescribed velocity $\V(s)$ and the rotlet-induced translation, the latter being $\bigO{\norm{\m}/\epsilon\eta}$ from \cref{eq: partially reduced constraint,eq: I2 exact}, so that
\begin{equation}
    \epsilon \norm{\u_{T0}} = \bigO{\epsilon\norm{\V}, \frac{\norm{\m}}{\eta}}\,.
\end{equation}
The size of the rotlet-generated angular velocity in \cref{eq: difference boundary condition} is more simply estimated as
\begin{equation}
    2\epsilon\eta(s)I_1(s)\m(s)\cross\er(s,\phi) = \bigO{\frac{\norm{\m}}{\epsilon\eta}}\,,
\end{equation}
recalling that $I_1(s)=\bigO{1/\epsilon^2\eta^2}$ from \cref{eq: I1 approx in middle}. With these estimates, we can see that the direct angular velocity contribution of the rotlets in \cref{eq: difference boundary condition} algebraically dominates the angular velocity induced by the force density if
\begin{equation}\label{eq: all conditions to ignore induced angular vel}
    \epsilon\norm{\V} = \bigO{\frac{\norm{\m}}{\eta}}\,.
\end{equation}
Should this condition hold, then we can neglect the term $\u_T(\X(s,\phi))-\u_T(\X(s,\phi+\pi))$ in \cref{eq: difference boundary condition} and incur $\bigO{\epsilon}$ relative errors. Indeed, if $\V(s)$ is smaller than the rotlet-induced translation, then this condition is satisfied without any additional constraints. Alternatively, if $\V(s)$ contributes to the dominant translational velocity, then our constraint becomes
\begin{equation}\label{eq: condition V m}
    \norm{\V} = \bigO{\frac{\norm{\m}}{\epsilon\eta}}\,,
\end{equation}
Further, we can estimate the magnitude of $\m$ by appealing to the calculations of \citet{Chwang1974}, or equivalently to those of \cref{sec: rotating}, to give the order-of-magnitude estimate $\norm{\m}=\bigO{\epsilon^2\eta^2\norm{\Omega}}$, which agrees with simple scaling arguments. Hence, \cref{eq: condition V m} becomes, after cancellation,
\begin{equation}\label{eq: bound on V and Omega}
    \norm{\V} = \bigO{\epsilon\eta\norm{\Omega}}\,.
\end{equation}
Thus, if the imposed translational velocity is at least a factor of $\epsilon$ smaller than the imposed angular velocity, we can neglect the force-induced angular velocity terms in \cref{eq: difference boundary condition} with $\bigO{\epsilon}$ relative errors.

In particular, we note that these conditions are satisfied precisely when the
magnitude of $\epsilon\Omega(s)$ is greater than or equal to that of $\V(s)$.
Hence, with reference to the no-slip condition of \cref{eq: general BC}, the
neglect of the force-induced angular velocity from \cref{eq: difference
boundary condition} is justified whenever the local rotation of the slender
body is making a leading-order contribution to the overall no-slip condition,
which is the focus and motivation of this study. In these circumstances, the
angular velocity boundary condition of \cref{eq: difference boundary
condition} reduces to
\begin{equation}\label{eq: general full}
    \Omega(s)\cross\er(s,\phi) = I_1(s)\m(s)\cross\er(s,\phi)
\end{equation}
with $\bigO{\epsilon}$ relative errors. Remarkably, this is precisely the relation of \cref{eq: almost rotating rrt}, valid whenever $\eta(s)\neq0$. Hence, following the reasoning of \cref{sec: rotating}, we once again arrive at the resistive torque theory relation
\begin{equation}\label{eq: general rtt}
    \Omega(s) = I_1(s)\m(s)\,,
\end{equation}
Thus, the results of \cref{sec: rotating} hold in more generality than we were
originally able to conclude, valid for all $\V(s)$ and $\Omega(s)$ satisfying
the above constraints. In particular, this resistive torque theory holds for
all components of $\Omega(s)$ and $\m(s)$, no longer limited to only axial
angular velocities and torques. This validity extends to the asymptotic
expansions of $I_1(s)$ both near-to and far-from the endpoints of the slender
body, so that, away from the ends of the body, we have the leading-order
result
\begin{equation}
    \frac{\epsilon^2\eta^2(s)}{2}\Omega(s) = \m(s)\,.
\end{equation}
Noting that the torque density is given by $8\pi\m (s)$, this slender-body
result is fully consistent with the calculation of \citet{Koens2018b} for the
torque density away from the end points of an axially rotating cylinder, which
here we restrict to be slender. Again with a slenderness restriction, we also
note that the above result is consistent with the calculation of the torque
density for an axially rotating torus \citep{Chwang1990,Thaokar2007}.

%% file: sections/numerics.tex
%!TEX root=../main.tex 

In this section, we will present a number of numerical examples to support our
asymptotic analysis. The numerical schemes used to explore these examples are
standard, making use of non-specialised quadrature routines and direct linear
solvers in MATLAB, with a comprehensive implementation available at
\url{https://gitlab.com/bjwalker/rotational-sbt}. Throughout, as introduced in
\cref{sec: invertibility}, we discretise the centreline of the slender body
between $-e$ and $e$ into $N$ equal intervals with midpoints $s_i\in(-e,e)$,
approximating unknown force and torque densities as constant on each segment.
We will typically take $N=400$ in the computations that follow.

Having previously discussed the numerical implementation of the rotlet ansatz
in \cref{sec: invertibility}, it remains to consider the discretised combined
ansatz of \cref{sec: translation}. In this case, with the discrete ansatz
having $2N$ vector unknowns, a natural choice would be to directly impose the
velocity boundary condition at $2N$ points on the surface, yielding a square
linear system. However, when implementing this in practice, we invariably
encountered linear systems with pathologically high condition numbers,
typically on the order of $10^9$, which posed a significant barrier to
accurate numerical solution. Though we are not aware of the precise origin of
this occurrence, we find that it may be successfully and simply circumvented
by imposing manipulated boundary conditions, reducing the condition number to
approximately $10^2$. We achieve this by once again making use of the partial
antisymmetry of the boundary condition on taking $\phi\mapsto\phi+\pi$, with
the details being presented in \cref{app: imposing combined ansatz BCs} and
following a similar principle to those detailed in \cref{sec: invertibility}
for the rotlet ansatz.

\subsection{Axially rotating slender bodies}\label{sec: axial rotation}
In order to evidence the ability of the pure rotlet ansatz of \cref{sec:
rotating} to capture the axial rotation of slender bodies with various
centrelines, we solve for the torque density in the rotlet ansatz as described
in \cref{sec: invertibility} for a range of slender bodies. The centrelines of
the considered slender bodies are illustrated in \cref{fig: axially rotating
bodies}, with explicit parameterisations given in \cref{app: centreline
params}. We uniformly take $\eta(s)=\sqrt{1-s^2}$, though the results that
follow are not sensitive to this choice. In order to quantify the accuracy
with which the rotlet ansatz satisfies the no-slip condition on the surface of
the slender body, we define the absolute error measure
\begin{equation}\label{eq: error def abs}
    \mathcal{E} = \max\limits_{s,\phi}\norm{\u - \u^{\text{num}}}_{\infty}\,,
\end{equation}
defining $\u=\u(\X(s,\phi))$ and $\u^{\text{num}}=\u^{\text{num}}(\X(s,\phi))$
to be the prescribed and numerically computed surface velocities at the point
$\X(s,\phi)$, respectively. This measure captures the greatest component-wise
discrepancy between the numerical solution and the prescribed velocity on the
surface, where the components are with respect to a fixed orthonormal
laboratory basis. 

\begin{figure}
    \centering
    \begin{tabular}{lccccc}
        &\begin{overpic}[width=0.17\textwidth]{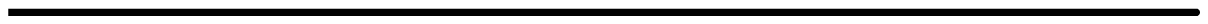}\put(0,90){(a)}\end{overpic} & \begin{overpic}[width=0.17\textwidth]{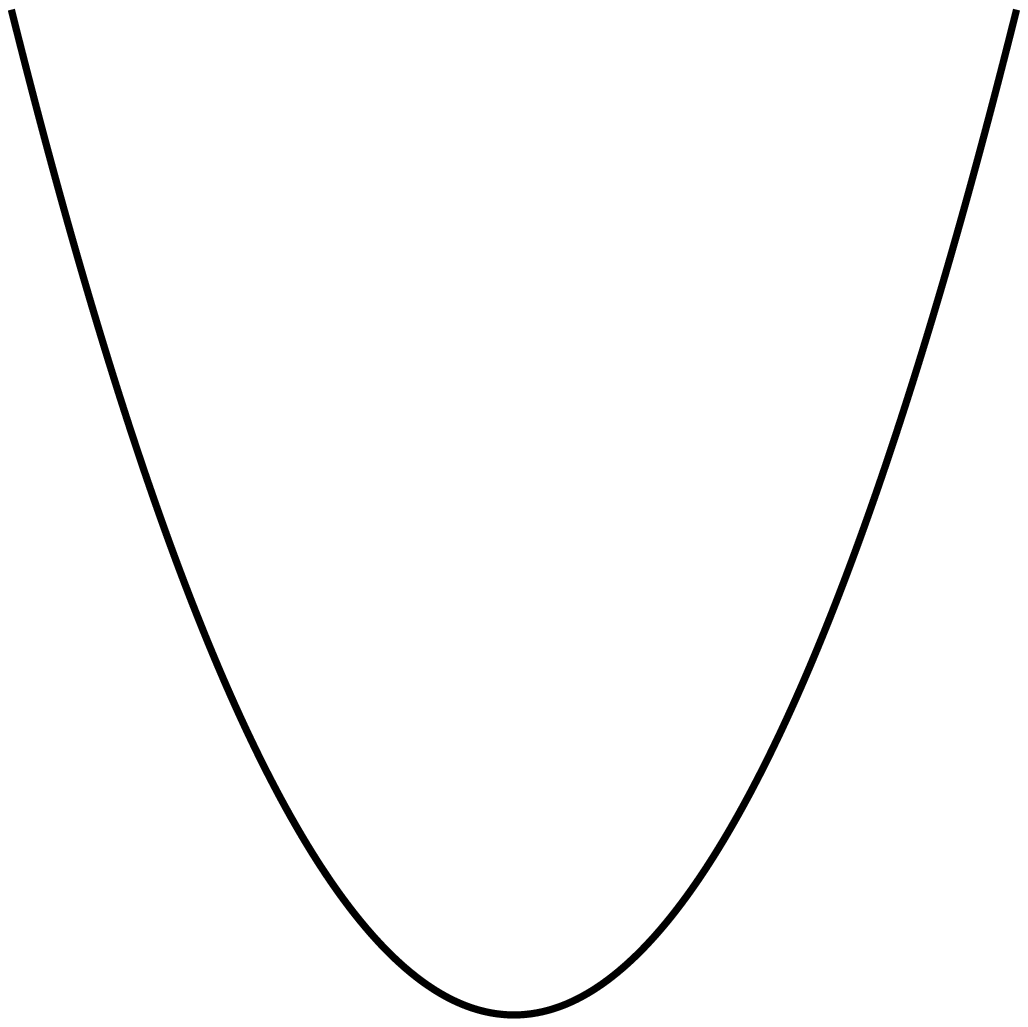}\put(0,90){(b)}\end{overpic} & \begin{overpic}[width=0.17\textwidth]{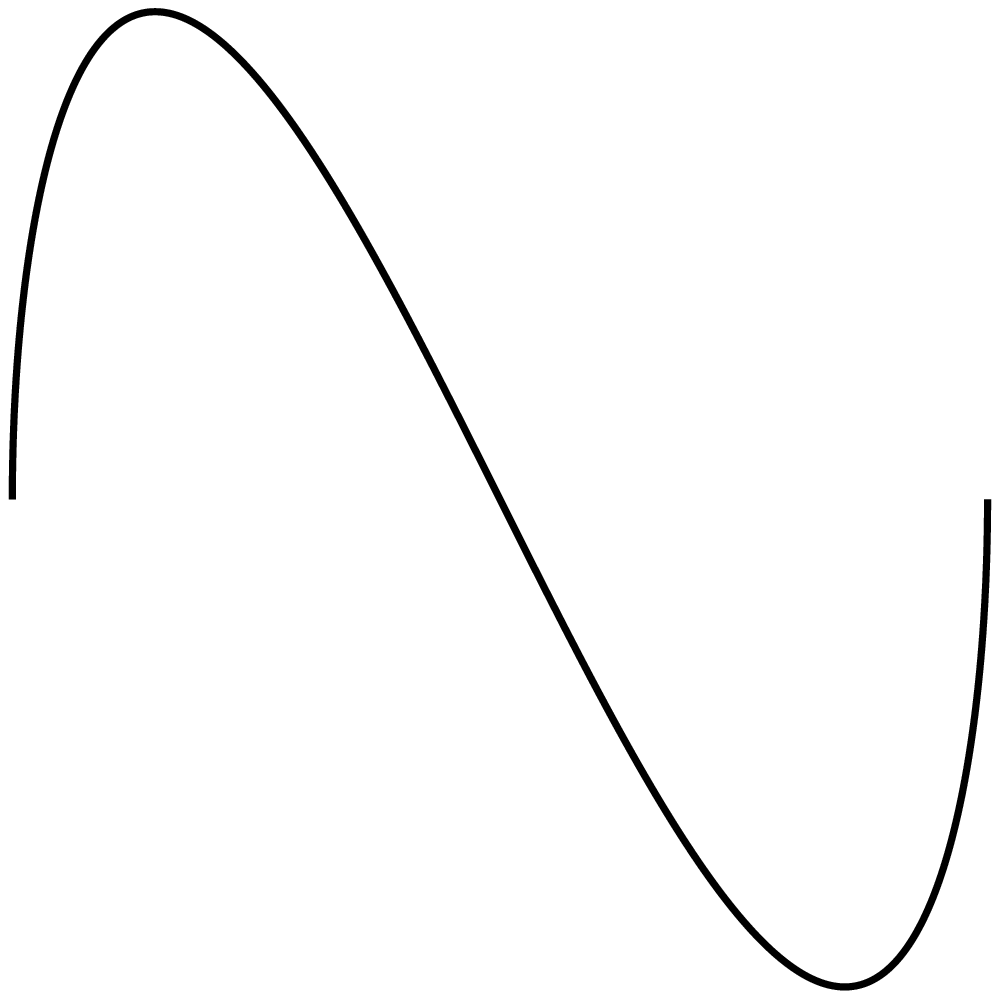}\put(0,90){(c)}\end{overpic} & \begin{overpic}[width=0.17\textwidth]{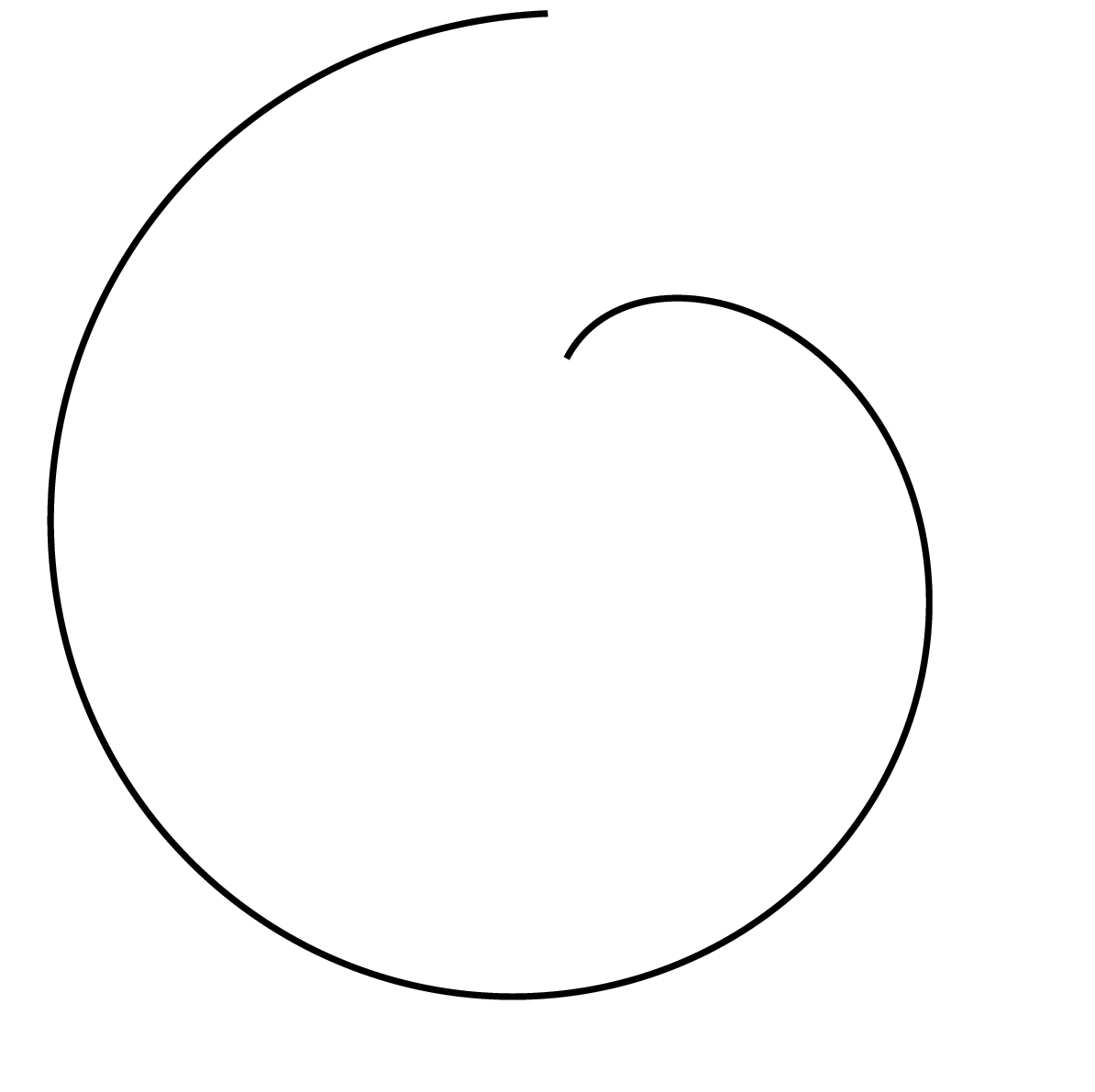}\put(0,90){(d)}\end{overpic} & \begin{overpic}[width=0.17\textwidth]{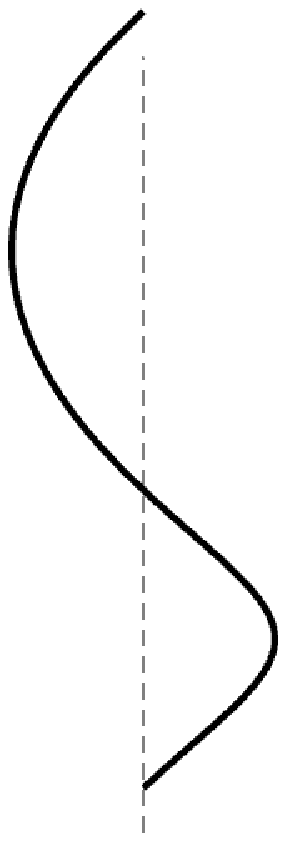}\put(0,90){(e)}\end{overpic}\\
        $\Erotlet_{\epsilon=0.01}$ & \num{8.13e-6} & \num{1.19e-3} & \num{1.55e-3} & \num{1.40e-3} & \num{7.11e-4}\\
        $\Erotlet_{\epsilon=0.005}$ & \num{1.64e-5} & \num{3.53e-4} & \num{4.71e-4} & \num{4.11e-4} & \num{2.38e-4}\\
        \hline
        $\ERTT_{\epsilon=0.01}$ & \num{6.12e-6} & \num{6.31e-4} & \num{8.59e-4} & \num{7.63e-4} & \num{3.88e-4}\\
        $\ERTT_{\epsilon=0.005}$ & \num{1.62e-5} & \num{1.83e-4} & \num{2.51e-4} & \num{2.16e-4} & \num{2.22e-4}\\
    \end{tabular}
    \caption{Maximum absolute error in the no-slip boundary condition for
    axially rotating slender bodies. We report the error $\mathcal{E}$ in the
    surface velocity generated by \cref{eq: pure rotlet ansatz}, making use of
    numerically computed torque densities that result from \cref{eq:
    rotational boundary condition ansatz} and \cref{eq: rotating rtt} in turn,
    denoting the corresponding errors by $\Erotlet$ and $\ERTT$, respectively.
    which correspond to the full rotlet boundary condition and the derived
    resistive torque theory, respectively. Having taken $\eta(s) =
    \sqrt{1-s^2}$ and discretising the centreline into $N=100$ segments, we
    report the maximum absolute errors, as measured in the infinity norm and
    defined in \cref{eq: error def abs}, for $\epsilon\in\{0.01,0.005\}$,
    prescribing $\V(s)=\vec{0}$ and $\Omega(s)=\et$. In (b--e), the
    numerically computed errors are seen to be scaling approximately with
    $\epsilon^2$, with
    $\mathcal{E}_{\epsilon=0.01}\approx4\mathcal{E}_{\epsilon=0.005}$, in line
    with the analysis of \cref{sec: rotating}, whilst the errors in (a) are
    dictated by the centreline discretisation and reduce upon refinement (see
    \cref{sec: verify prolate spheroid}). The slender bodies with the largest
    errors correspond to those with the largest curvatures, with the largest
    errors being localised to regions with higher curvature. These plots have
    aspect ratio 1:1 and are independently scaled for visual clarity.
    Parameterisations of these centrelines are given in \cref{app: centreline
    params}.}
    \label{fig: axially rotating bodies}
\end{figure}

For $\epsilon\in\{0.01,0.005\}$, a discrete approximation to this error is
reported for both the full rotlet ansatz of \cref{eq: rotational boundary
condition ansatz}, denoted by $\Erotlet{}$, and the resistive torque theory
result of \cref{eq: rotating rtt}, denoted by $\ERTT{}$. In both cases, the
computed $\m(s)$ were inserted into \cref{eq: pure rotlet ansatz}, which was
then evaluated at 612 distinct points on the surface of the body, at 51
discrete arclengths and 12 equispaced points around the circumference of each
discrete cross section. In each of the cases reported, the rotlet ansatz can
be seen to accurately capture the surface velocity of the slender body, with
errors approximately on the order of $\epsilon^2$, as predicted by the
analysis of \cref{sec: rotating}. The scaling of the error with $\epsilon^2$
is confirmed by comparing the errors for the two considered values of
$\epsilon$, denoted via subscripts as $\mathcal{E}_{\epsilon = 0.01}$ and
$\mathcal{E}_{\epsilon = 0.005}$, noting that
$\mathcal{E}_{\epsilon=0.01}\approx4\mathcal{E}_{\epsilon=0.005}$ in all cases
except for the prolate spheroid in \cref{fig: axially rotating bodies}a. The
error corresponding to the prolate spheroid is significantly lower than the
other reported slender bodies, instead being limited by the accuracy of the
employed quadrature and the discretisation of the slender body. We will return
to consider the prolate spheroid, and the increased accuracy of the ansatz in
this case, in \cref{sec: verify prolate spheroid}. Remarkably, in all cases,
the errors associated with the resistive torque theory result of \cref{eq:
rotating rtt} are of the same order as those associated with the full ansatz,
including those slender bodies with curved centrelines, evidencing the
validity of utilising the resistive torque theory in calculations involving
axially rotating slender bodies.

\subsection{Combining translation and rotation}
\begin{figure}
    \centering
    \begin{overpic}[permil,width=0.8\textwidth]{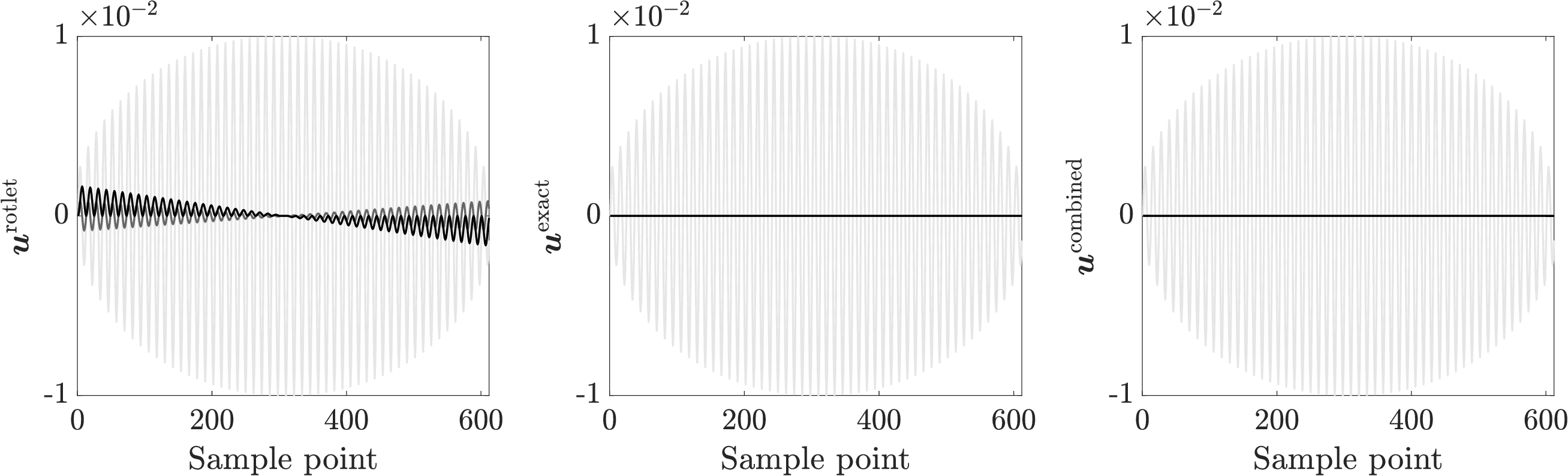}
        \put(-4,290){(a)}
        \put(335,290){(b)}
        \put(677,290){(c)}
        % \put(-4,200){(d)}
        % \put(335,200){(e)}
        % \put(677,200){(f)}
    \end{overpic}
    \caption{Off-axis rotation of a prolate ellipsoid. Imposing the angular
    velocity $\Omega(s) = \en(s)$ on a straight prolate ellipsoid, taking
    $\V(s)=\vec{0}$ and $\epsilon = 0.01$, we numerically solve for the forces
    and/or torques in the pure rotlet of \cref{sec: rotating} and the combined
    ansatz of \cref{sec: translation}. We sample the induced velocity at 612
    points on the surface of the slender body, sampling at 12 equispaced
    angles at 51 equally spaced arclengths. The components of velocity at the
    sample points are shown for the rotlet ansatz, the prescribed boundary
    condition, and the combined ansatz in panels (a), (b), and (c),
    respectively, with different colours corresponding to different
    components. The $\et$ component, shown lightest, is captured well by both
    ansatzes, whilst the $\en$ and $\eb$ components, shown as dark grey and
    black curves, respectively, deviate significantly from the exact solution
    when using the pure rotlet ansatz. The combined ansatz of \cref{sec:
    translation} accurately captures all velocity components, with errors on
    the order of $10^{-6}$. Sample points are in ascending order by arclength,
    with perceived oscillations corresponding to sampling at different points
    around the circumference of each cross section.}
    \label{fig: offaxis rotation}
\end{figure}

Having established that the pure rotlet ansatz is capable of accurately
satisfying velocity boundary conditions for axially rotating slender bodies,
we now numerically explore its efficacy for capturing off-axis rotation of a
prolate ellipsoid with $\epsilon=0.01$, comparing its performance against the
combined ansatz of \cref{sec: translation}. Imposing the off-axis angular
velocity $\Omega(s) = \en(s)$ and taking $\V(s)=\vec{0}$, we numerically solve
for the forces and/or torques in the two ansatzes and evaluate the velocity at
the surface evaluation points described in \cref{sec: axial rotation}. The
components of this velocity, including the prescribed boundary condition for
comparison, are shown in \cref{fig: offaxis rotation}, denoted
$\u^{\text{rotlet}}$, $\u^{\text{combined}}$, and $\u^{\text{exact}}$,
respectively. As predicted by the analysis of \cref{sec: rotating}, the rotlet
ansatz is unable to capture the rotation without introducing additional,
erroneous components of velocity on the order of $10^{-3}$ in this case,
whilst the combined ansatz successfully compensates for the rotlet-induced
velocity and satisfies the boundary condition with $\mathcal{E} \approx
10^{-6}$.

\subsection{Neglecting the force-induced couple}
In \cref{sec: generalised resistive torque theory}, we identified a regime in
which the generalised resistive torque theory was accurate to leading order,
which corresponded to cases where the angular velocity contributes to the
velocity boundary condition at leading order. To numerically verify this
conclusion, we consider a parabolic slender body, with centreline
corresponding to \cref{fig: axially rotating bodies}b, taking
$\eta(s)=\sqrt{1-s^2}$ and $\epsilon=0.01$. For this slender body, we solve
for the forces and torques on the body with $\V(s)=\ex$, where
$\{\ex,\ey,\ez\}$ are an orthonormal basis of the laboratory frame, and
$\Omega(s) \parallel \et$, varying the magnitude of $\Omega(s)$. Here, we
discretise the slender body into $N=100$ segments.

\begin{figure}
    \centering
    \includegraphics[width = 0.4\textwidth]{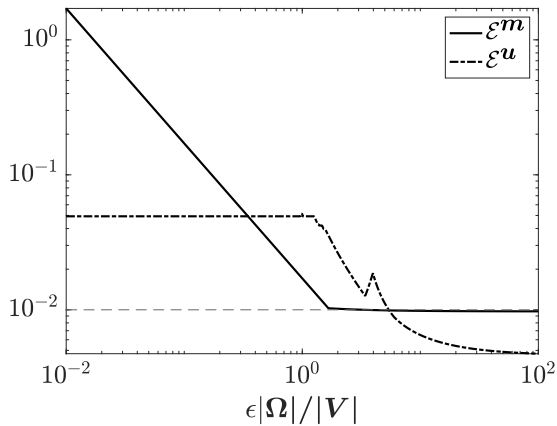}
    \caption{Relative error in using the generalised resistive torque
    theory. Numerically computing the torques associated with the resistive
    torque theory relation of \cref{eq: general rtt}, we report the maximum
    relative error in torque when using the resistive torque theory, denoted
    by $\mathcal{E}^{\m}$. This is measured relative to the torque computed
    from enforcing the full boundary condition of \cref{eq: combined ansatz
    full bc}, and is shown as a solid curve and as a function of the ratio
    $\epsilon\norm{\Omega}/\norm{\V}$. The analogous error in the surface
    velocity, denoted by $\mathcal{E}^{\u}$ and evaluated at the discrete
    surface points described in \cref{sec: axial rotation}, is shown
    dot-dashed. In line with the analysis of \cref{sec: generalised resistive
    torque theory}, the errors in both the torque and the surface velocity are
    approximately on the order of $\epsilon$, here $10^{-2}$, when
    $\epsilon\norm{\Omega}/\norm{\V}>1$. The apparent spike in
    $\mathcal{E}^{\u}$ is an artefact of the discrete approximation to the
    maximum relative error. Here, we have taken $N=100$, $\V(s)=\ex$ and
    $\Omega(s)\parallel \et$, adopting the parabolic centreline of \cref{fig:
    axially rotating bodies}b and fixing $\eta(s)=\sqrt{1-s^2}$.}
    \label{fig: validity of RTT}
\end{figure}

In \cref{fig: validity of RTT}, we show the maximum relative difference
between the torques computed by the combined ansatz of \cref{eq: combined
ansatz full bc} and the resistive torque theory of \cref{sec: translation},
treating the combined ansatz as the gold standard in the absence of an
analytical solution. In symbols, this relative error measure
$\mathcal{E}^{\m}$ is defined as
\begin{equation}
    \mathcal{E}^{\m} = \max\limits_{s,\phi}\frac{\norm{\m^{\text{RTT}} - \m^{\text{combined}}}_{\infty}}{\norm{\m^{\text{combined}}}_{\infty}}\,,
\end{equation}
where $\m^{\text{combined}}$ and $\m^{\text{RTT}}$ are the torques computed
from the combined ansatz and from the resistive-torque-theory approximation to
the combined ansatz, respectively. Note that, in the resistive torque theory
solution, we directly compute the torques from the imposed angular velocity,
then impose the full boundary condition of \cref{eq: combined ansatz full bc}
at $N$ points on the surface of the body to yield the force density. Alongside
this error, we show the analogous relative error in the velocity, defined as
\begin{equation}
    \mathcal{E}^{\u} = \max\limits_{s,\phi}\frac{\norm{\u^{\text{RTT}} - \u^{\text{combined}}}_{\infty}}{\norm{\u^{\text{combined}}}_{\infty}}\,,
\end{equation}
where $\u^{\text{combined}}$ and $\u^{\text{RTT}}$ are the velocities
corresponding to the two ansatzes, respectively, generated by evaluating
\cref{eq: combined ansatz} with the computed forces and torques. These
errors are shown as functions of the ratio between the magnitude of the
angular velocity and the magnitude of the translational velocity, scaled by
$\epsilon$, with reference to the condition of \cref{eq: bound on V and
Omega}. In particular, having taken $\epsilon=0.01$ in this example, the
relative error in both the torque and velocity is on the order of $\epsilon$
for $\norm{\V} = \epsilon\norm{\Omega}$, as predicted by the asymptotic
analysis of \cref{sec: generalised resistive torque theory}.

\subsection{Verification against an exact solution}\label{sec: verify prolate spheroid}
Finally, as a limited yet informative analytical verification, we compare the computed forces and torques corresponding to the combined ansatz of \cref{sec: translation} with the exact solution for the simultaneous rigid body translation and axial rotation of a prolate spheroid, as presented by \citet{Chwang1974,Chwang1975}. Specifically, we let $\et = \ex$ and prescribe $\V(s) = (\ex + \ey + \ez) / \sqrt{3}$ and $\Omega(s) = \et$. \Citeauthor{Chwang1974}'s results yield the exact relations
\begin{subequations}
\begin{align}
    \f(s) &= \left[\frac{e^2}{-2e + (1+e^2)\log\frac{1+e}{1-e}}\et\otimes\et +  \frac{2e^2}{2e + (3e^2-1)\log\frac{1+e}{1-e}}\left(\eye - \et\otimes\et\right) \right] \V\,,\\
    \m(s) &= \frac{e^2(1-s^2)}{\frac{2e}{1-e^2} - \log\frac{1+e}{1-e}} \Omega\,,\label{eq: CW exact sol torque}
\end{align}
\end{subequations}
which in fact hold for any constant translation $\V$ and constant axial
rotation $\Omega$. On writing this, we are setting $\eta(s) = \sqrt{1-s^2}$,
which gives rise to the parabolic torque profile of \cref{eq: CW exact sol
torque}. We compute the analogous numerical solutions for $\f$ and $\m$ from
the full ansatz of \cref{eq: combined ansatz full bc}, as described in
\cref{app: imposing combined ansatz BCs}, discretising the slender body using
$N=400$ segments and taking $\epsilon=0.01$. We observe componentwise errors
on the order of $10^{-9}$ in the forces and torques, limited by the numerical
precision of the quadrature employed and the level of discretisation of the
slender body. When employing the resistive torque theory of \cref{eq: general
rtt}, these errors remain small but increase to approximately $10^{-6}$ and
$10^{-8}$ for the force and torque, respectively, which further highlights the
validity of the leading-order resistive-torque-theory approximation.

%% file: sections/discussion.tex
%!TEX root=../main.tex 

In this study, we have posed and explored a simple rotlet ansatz for the flow
around a locally rotating slender body, motivated by the classical solutions
of \citet{Chwang1974} for axisymmetric bodies. In doing so, we have seen how
such a rotlet ansatz is capable of capturing surface flows that result from
locally axial rotations, whilst it can induce unwanted translational
velocities if slender-body rotation is not purely axial. These induced flows
were found to satisfy volume conservation to leading asymptotic order and,
hence, we were able to make use of an established slender-body theory for
translational motion to compensate for the linear velocities
\citep{Walker2020b}. This resulted in a combined theory that is uniformly accurate to
leading algebraic order in the aspect ratio of the slender body. Our combined
ansatz is broadly applicable, being suited to bodies with circular cross
sections and otherwise general shapes, including curved centrelines and
spatially varying cross-sectional radii, subject to reasonable constraints on
the centreline curvature and the arclength derivative of the radius function.
Hence, we envisage its use in the future theoretical study of a range of
slender bodies, in particular biological filaments such as cilia and flagella.
Previous exploration into these application areas has made use of ansatzes
similar to that developed here \citep{Ishimoto2018a,Carichino2019,Huang2019},
though a detailed analysis of the suitability of the various ansatzes has been
absent from previous works, to the best of our knowledge. Here, we have
justified the validity of our posed ansatzes through a formal asymptotic
analysis, ascertaining the capabilities and limitations of a pure rotlet
ansatz and the flexibility afforded by a combined theory.

A notable consequence of our asymptotic analysis of the rotlet ansatz was the
emergence of a local, leading-order relation between the angular velocity and
torque, reminiscent of the formula provided by \citet{Chwang1974} for the
torque on a rotating circular cylinder and the uniaxial result of \citet{Keller1976}. This so-called resistive torque
theory, initially shown to be valid for axial rotations then extended to apply
to general slender bodies (appropriately qualified in \cref{sec: generalised
resistive torque theory}), arose with relative simplicity compared to the
canonical resistive force theories developed since the mid 20th century
\citep{Gray1955,Hancock1953,lighthill1976}. The ready formulation of this
theory was a consequence of the sharp near-singular nature of the rotlet
kernel, which results in small region of the domain of integration being the
algebraically dominant contribution to the rotlet ansatz. In contrast, the
analogous Stokeslet integral kernel is less sharp, giving rise to the
logarithmic errors typically associated with resistive force theories
\citep{Johnson1980}. 

Owing to the algebraic errors and marked simplicity of the presented resistive
torque theory, together with the absence of severe constraints, we anticipate
that it will be of broad utility to slender body hydrodynamics. In particular,
the combined theory of this study contains no restrictions preventing its
application to chiral filaments, for instance, nor to the mobility problem,
where velocities and angular velocities are determined given known force and
torque densities. Furthermore, the presented framework provides an intuitive,
leading-order link between angular velocity and torque that aligns with the
classical results of \citet{Chwang1974}, with simple ansatzae that are readily
recognised as generalisations of this classical study.  Finally, the present
analysis also serves to justify the recent use of \citeauthor{Chwang1974}'s
relation in the elastohydrodynamic method of \citet{Walker2020a}, though
\citeauthor{Walker2020a}'s work does not include non-local translational
hydrodynamics.

When evaluating our slender-body ansatzes numerically in \cref{sec: numerics},
we found that isolating the translational and angular velocities in turn
significantly improved the condition number of the associated linear systems,
which we achieved by taking various linear combinations of the boundary
conditions. However, this result is purely empirical, with detailed analysis
of this phenomenon warranting future consideration. In the case of the rotlet
ansatz, we leveraged the same idea of taking particular linear combinations
when forming our system of equations, which then enabled us to guarantee the
invertibility of the system subject to the approximate bound $\epsilon N < 1$.
Though this condition is merely sufficient, not necessary, for invertibility,
it suggests that the discretised ansatz may be unsuitable for numerical
solution in this way if $N$ is larger than $1/\epsilon$. This observation
aligns qualitatively with anecdotal reports of instabilities identified in
other slender-body theories, with $N\approx 1/\epsilon$ often resulting in
poor conditioning, or even numerical singularity, in practice. To the best of
our knowledge, this issue remains both unreported and uninvestigated in the
literature, with exploring the practical invertibility of the linear systems
associated with discretised slender-body theories thereby representing a
pertinent topic for future work, particularly given the popularity of
slender-body theories. Another topic for future investigation is the
accommodation of rotation in slender-body theories when the rotation is not
sufficiently large for it to be within the leading-order kinematics, noting
that recent advances in determining the higher order of slender-body theory by
\citet{Koens2022}, albeit possibly with a greater degree of non-locality.

In summary, we have explored the hydrodynamics of slender bodies in the Stokes
regime, explicitly seeking to capture their local rotation via a slender-body
ansatz for the fluid flow. Through the analysis of a classically inspired line
distribution of rotlets, we have posed and validated a combined ansatz for
capturing both translation and local rotation with errors that are uniformly algebraic
in the slender body's aspect ratio. Our asymptotic analysis also revealed a
surprisingly simple local relation between the angular velocity and the torque
on the body, leading to an algebraically accurate resistive torque theory that
complements classical resistive force theories.

%% file: sections/appendices.tex
%!TEX root=../main.tex
\section{Estimating  the volume flux rate integral}
\label{app: vol flux}
For an arclength-dependent vector field $\vel(s)$ we estimate the magnitude of the net volume flux rate integral
\begin{equation}\label{eq: app: net volume flux def}
    \dVol (\vel) \coloneqq \iint \vel(s)\cdot\intd{\vec{S}} = \int_{-e}^e \int_0^{2\pi} \vel(s)\cdot \left(\pdiff{\X}{\phi}\cross\pdiff{\X}{s}\right)\intd{\phi}\intd{s} \,,
\end{equation}
where $\X(s,\phi)$ is as defined in \cref{eq:surfacepoint}, the surface normal is outward and  the integration is over the entire surface of the slender body. In order to evaluate the integral in \cref{eq: app: net volume flux def}, we compute the partial derivatives of $\X(s,\phi)$ with respect to $s$ and $\phi$, yielding
\begin{subequations}
\begin{align}
    \pdiff{\X}{s} &= \left[1 - \epsilon\eta\kappa\cos{\phi}\right]\et + \epsilon\eta'\er + \epsilon\eta\tau\ephi\,, \label{eq: derivs of X: s}\\
    \pdiff{\X}{\phi} &= \epsilon\eta\ephi\,,\label{eq: derivs of X: phi}
\end{align}
\end{subequations}
where $\eta'$ denotes $\mathrm{d}\eta/\mathrm{d}s$ and we define the azimuthal unit vector
\begin{equation}
    \ephi(s,\phi) \coloneqq \pdiff{\er}{\phi} = -\sin{\phi}\en + \cos{\phi}\eb\,,
\end{equation}
which can equivalently be defined as $\ephi\coloneqq\et\cross\er$\,. To arrive at \cref{eq: derivs of X: s}, we have made use of the additional Frenet-Serret relations
\begin{equation}
    \pdiff{\en}{s} = -\kappa\et + \tau\eb\,, \quad \pdiff{\eb}{s} = -\tau\en\,,
\end{equation}
in which $\kappa$ is the centreline curvature and $\tau$ is the accompanying torsion, the latter of which quantifies the non-planarity or intrinsic twist of the centreline. \Cref{eq: derivs of X: s,eq: derivs of X: phi} then lead to
\begin{equation}\label{eq: normal and surface element}
    \pdiff{\X}{\phi}\cross\pdiff{\X}{s} = \left[\epsilon\eta - \epsilon^2\kappa\eta^2\cos{\phi}\right]\er - \epsilon^2\eta\eta'\et\,,
\end{equation}
noting that $\ephi\cross\et=\er$ and $\ephi\cross\er=-\et$. In particular, \cref{eq: normal and surface element} captures the intuitive notion that the local surface normal is aligned approximately with $\er$, with corrections proportional to the rate of change of the cross-sectional radius with arclength. 

Returning to \cref{eq: app: net volume flux def}, we have
\begin{subequations}\label{eq: app: dVol}
\begin{align}
    \dVol(\vel) &= \epsilon\int_{-e}^e \eta(s)\vel(s)\cdot\int_0^{2\pi} \left[1 - \epsilon\kappa\eta(s)\cos{\phi}\right]\er(s,\phi)\intd{\phi}\intd{s}
    \,,\\ &+  
    2\pi \epsilon^2\int_{-e}^e \eta(s)\eta'(s) \vel(s)\cdot \bm e_t(s)  \intd{s} \,,\\
    &= 2\pi\epsilon^2\left[ \bigO{\kappa\sup_{s}[\eta^2\norm{\vel}]}+ \bigO{\sup_{s}[|\eta\eta'|\norm{\vel\cdot\bm e_t}]}\right] 
\end{align}
\end{subequations}
as $\int_0^{2\pi}\er\intd{\phi}=\vec{0}$ and $\int_0^{2\pi}\cos{\phi}\er\intd{\phi}=\pi\en$ by the definition of $\er$ in \cref{eq: er definition}.

\section{Numerical implementation of the combined ansatz}\label{app: imposing combined ansatz BCs}
Having discretised the slender body into $N$ segments as described in
\cref{sec: numerics}, naively enforcing the velocity boundary condition of
\cref{eq: combined ansatz full bc} at $2N$ points on the surface of the
slender body yields linear systems that, whilst empirically found to be
invertible, typically have very high condition numbers, on the order of $10^9$
for $\epsilon = 0.01$ and $N=100$. However, we have found that enforcing
alternative, mathematically equivalent conditions circumvents this problem,
though we lack rigorous justification of this resolution. In detail, drawing
inspiration from the invertible linear system constructed in \cref{sec:
invertibility} and the approach of \cref{sec: enforcing no-slip}, we impose
six scalar boundary conditions at a fixed $s=s_i$ on the
$i$\textsuperscript{th} segment. To do so succinctly, we note that the
boundary condition of \cref{eq: combined ansatz full bc} can be written as
\begin{equation}
    \u(\X(s,\phi)) = \u_T(\X(s,\phi)) + \u_R(\X(s,\phi))\,.
\end{equation}
We further simplify this notation in this context by fixing $s=s_i$, so that
$\u$, $\u_T$, and $\u_R$ depend only on $\phi$, and retain only the explicit
$\phi$ dependence of these terms, writing the boundary condition as
\begin{equation}
    \u(\phi) = \u_T(\phi) + \u_R(\phi)
\end{equation}
for brevity. With this compact notation, the first three scalar conditions that we impose involve taking the difference between the original boundary conditions evaluated at angles $\phi$ and $\phi+\pi$, as we did in \cref{sec: enforcing no-slip}, then projecting onto the local basis, as in \cref{sec: invertibility}. They read
\begin{subequations}
    \begin{align}
        [\u(0) - \u(\pi)] \cdot \et &= \left([\u_T(0) - \u_T(\pi)] + [\u_R(0) - \u_R(\pi)]\right)\cdot \et\,,\\
        [\u(0) - \u(\pi)] \cdot \eb &= \left([\u_T(0) - \u_T(\pi)] + [\u_R(0) - \u_R(\pi)]\right)\cdot \eb\,,\\
        [\u(\pi/2) - \u(3\pi/2)] \cdot \et &= \left([\u_T(\pi/2) - \u_T(3\pi/2)] + [\u_R(\pi/2) - \u_R(3\pi/2)]\right)\cdot \et\,,
    \end{align}
\end{subequations}
analogous to those of \cref{eq: rotlet bc projection} and implicitly
evaluating $\et$ and $\eb$ at $s=s_i$. These conditions seek to capture only
angular velocity contributions, with all $\phi$-independent terms cancelling
upon subtraction, as described in \cref{sec: enforcing no-slip}. Notably,
considering the leading-order terms in these boundary conditions leads to
equations similar to that of \cref{eq: difference boundary condition}, with
the choice of components here guaranteeing invertibility of the rotlet
contribution as in \cref{sec: invertibility}. We impose the three remaining
scalar conditions by following a similar principle, but instead consider sums
of the surface velocities, eliminating angular terms to isolate the effects of
translation. These conditions can be summarised by the single vector condition
\begin{equation}
        [\u(0) + \u(\pi)] = [\u_T(0) + \u_T(\pi)] + [\u_R(0) + \u_R(\pi)]\,.
\end{equation}
This overall approach can be viewed as attempting to decouple the
translational and rotational flow problems. Empirically, we have found that
this leads to linear systems of unproblematic condition number, typically on
the order of $10^2$ for $\epsilon=0.01$ and $N=100$, though establishing this
analytically remains a topic for future study.

\section{Centreline parameterisations}\label{app: centreline params}
The centrelines of \cref{fig: axially rotating bodies} may be parameterised in Cartesian $xyz$-coordinates as
\begin{equation*}\begin{array}{lll}
    \text{(a)} & (x(t),y(t),z(t))=(t,0,0)\,, & t\in[-1,1]\,,\\
    \text{(b)} & (x(t),y(t),z(t))=(t,t^2,0)\,, & t\in[-1,1]\,,\\
    \text{(c)} & (x(t),y(t),z(t))=(\sin{t},\sin{t}\cos{t},0)\,, & t\in[\pi/2,3\pi/2]\,,\\
    \text{(d)} & (x(t),y(t),z(t))=(t^{1/4}\sin{t},t^{1/4}\cos{t},0)\,, & t\in[0.1,2\pi]\,,\\
    \text{(e)} & (x(t),y(t),z(t))=(\sin{t},\cos{t},t)\,, & t\in[0,2\pi]\,,
    \end{array}
\end{equation*}
where labels correspond to those in \cref{fig: axially rotating bodies}. Arclength parameterisations of these curves are computed numerically
to obtain $\xi(s)$.